\documentclass[11pt,a4paper]{article}
\usepackage{jheppub}

\usepackage{float}
\usepackage{empheq}
\usepackage{multirow}
\usepackage{cleveref}
	\crefname{ineq}{ineq.}{ineqs.}
	\Crefname{ineq}{Ineq.}{Ineqs.}
	\creflabelformat{ineq}{\upshape(#2#1#3)}
	\crefname{stat}{statement}{statements}
	\Crefname{stat}{Statement}{Statements}
	\creflabelformat{stat}{\upshape(#2#1#3)}
	\crefname{cond}{condition}{conditions}
	\Crefname{cond}{Condition}{Conditions}
	\creflabelformat{cond}{\upshape(#2#1#3)}
\usepackage{slashed}
\usepackage{nicefrac}
\usepackage[percent]{overpic}
\usepackage[export]{adjustbox}	
\usepackage{tikz}
\usetikzlibrary{arrows, decorations.pathmorphing}

\usepackage[ddmmyyyy]{datetime}

\newcommand{\gev}{\text{ GeV}}

\numberwithin{equation}{section}
\allowdisplaybreaks

\def\beq#1\eeq{\begin{align}\begin{aligned} #1 \end{aligned}\end{align}}
\def\beqsep#1\eeqsep{\begin{align} #1 \end{align}}
\def\bbeq#1\ebeq{\begin{empheq}[box=\fbox]{align}\begin{aligned} \quad #1 \end{aligned}\end{empheq}}
\def\beg#1\eeg{\begin{gather}\begin{gathered} #1 \end{gathered}\end{gather}}

\makeatletter
\newcommand{\leqnomode}{\tagsleft@true}
\newcommand{\reqnomode}{\tagsleft@false}
\makeatother

\renewcommand{\phi}{\varphi}

\newcommand{\sgn}{\text{sgn}}
\newcommand{\eff}{\text{eff}}
\newcommand{\sqrts}{\sqrt{s}}
\newcommand{\eps}{\varepsilon}
\newcommand{\sym}{\text{sym}}
\renewcommand{\sl}{\slashed}
\newcommand{\tr}{\text{\,tr\,}}
\newcommand{\im}{\Im}
\renewcommand{\vec}[1]{{\bf #1}}
\newcommand{\n}[1]{|\vec #1|}
\newcommand{\cm}{\text{cm}}
\newcommand{\topstr}{\texorpdfstring}

\newcommand{\ca}{a}
\newcommand{\cb}{b}

\title{Thermal regularization of $t$-channel singularities in cosmology and particle physics:	the general case}

\author{Micha{\l} Iglicki}

\affiliation
{
	Faculty of Physics, University of Warsaw,\\
	ul. Pasteura 5, 02-093 Warsaw, Poland
}

\emailAdd{michal.iglicki@fuw.edu.pl}

\abstract
{
	This paper presents a~way to regularize the $t$-channel singularity (which appears when a~massive, stable $t$-channel mediator of a~given process is allowed to be on-shell, making the cross section infinite) in a~general case of particles of any spin (0,~\nicefrac12,~1) interacting within a~thermal medium. Those interactions result in a~finite lifetime of the mediator and allow to introduce an~effective momentum- and temperature-dependent width. As a~result, the would-be-singular cross section becomes finite. A~complete derivation and an~analytical result for the width are provided. For an~illustration, the method is used to calculate the thermal widths and cross sections within the~Vector-Fermion Dark Matter model.
}


\arxivnumber{2212.00561}

\begin{document}

\maketitle

\section{Introduction}
\label{sec:intro}
\subsection{\topstr{$t$}{t}-channel singularity and its relevance}
A~$t$-channel singularity of a~given scattering process with a~$t$-channel mediator arises when the mediator is kinematically allowed to be on its mass-shell. Then, the mediator's propagator becomes singular. For a~massive mediator, one cannot use the infrared regularization schemes. If, in addition, the mediator is stable, the usual Breit-Wigner approach cannot be applied to regularize the singularity using the mediator's width. This leads to a~truly singular (infinite) cross section.

Examples of Standard Model processes affected by the singularity include the weak analogue of the Compton scattering, $Ze^-\to e^-Z$, mediated by electron neutrino, and neutrino-mediated muon-muon scattering, $\mu^+\mu^-\to W^+W^{-*}\to W^+ e^-\bar\nu_e$. However, the most natural context in which the $t$-channel singularity appears are models of dark matter, as they provide massive stable particles that can serve as singular mediators.

It should be stressed that the $t$-channel singularity is a~serious issue whose role cannot be reduced to some kind of higher-order corrections to nevertheless finite results. Under certain circumstances, the discussed singularity makes cross sections truly infinite. Therefore, an~applicable and practical solution to this problem should be desired.

\subsection{Known approaches}
The existence of processes with a~$t$-channel mediator kinematically allowed to be on-shell is known at least since early '60s~\cite{Peierls:1961zz}. In 1965, Coleman and Norton \cite{Coleman:1965xm} proved that \emph{a~Feynman amplitude has singularities on the physical boundary if and only if the relevant Feynman diagram can be interpreted as a~picture of an~energy- and momentum-conserving process occurring in space-time, with all internal particles real, on the mass shell, and moving forward in time}, which in the case of a~$2\to 2$ $t$-channel process is actually equivalent to the~\cref{cond:sing} formulated here. Since then, however, the topic has not been widely explored by particle physicists apart of a~few papers (see, e.g., \cite{Brayshaw:1978xt}).

Dating from the '90s, several studies \cite{Ginzburg:1995bc,Melnikov:1996na,Melnikov:1996iu,Melnikov:1996ft,Dams:2002uy,Dams:2003gn} concerning the~$t$-channel singular process of $\mu^+\mu^-\to W^+e^-\bar\nu_e$ appeared due to its relevance for planned lepton colliders. The authors mainly proposed to cure the~singularity by including corrections resulting from a~finite size of the scattering beams. None of their proposals, however, is simultaneously fully reliable and applicable in the case of singular processes in the early Universe, when no beams are involved so there is no scale to be used as a~regulator.

A similar mechanism has been introduced in \cite{Ioannisian:1998ch} and further developed in \cite{Karamitros:2022fki}. The authors encountered the problem of $t$-channel singularity while considering neutrino oscillations. They propose to regularize the singularity by taking into account non-locality of interactions within the source and the detector. Although promising, that approach cannot be applied to the case this paper focuses on, for the reasons similar to those mentioned above.

Another natural idea is to include thermal corrections to the masses of the particles involved in the process, hoping that the singularity disappears, as it happens with the IR singularities in the HTL approach \cite{Braaten:1989mz}. Unfortunately, for a massive mediator, this would only shift the singular value of momentum squared from the bare to the thermal mass squared. This could regularize the singularity only if the masses change dramatically, so that the decomposition corresponding to \cref{cond:sing} is no longer possible.

On the other hand, for cosmological applications, a~quite different method has been developed by the authors of \cite{Giudice:2003jh} inspired by \cite{Weldon:1983jn,Dicus:1982bz}. Their method bases on carefully taking into account the statistical factors present in the Boltzmann equation term corresponding to the singular process. Note that in his paper \cite{Weldon:1983jn}, Weldon obtained a~result similar to \cref{eq:selfEFinal}, working in the imaginary-time (Matsubara) formalism. His result agrees with the one presented in this paper, calculated within the real-time approach, but differs by a~numerical factor, see \cref{app:weldon}.

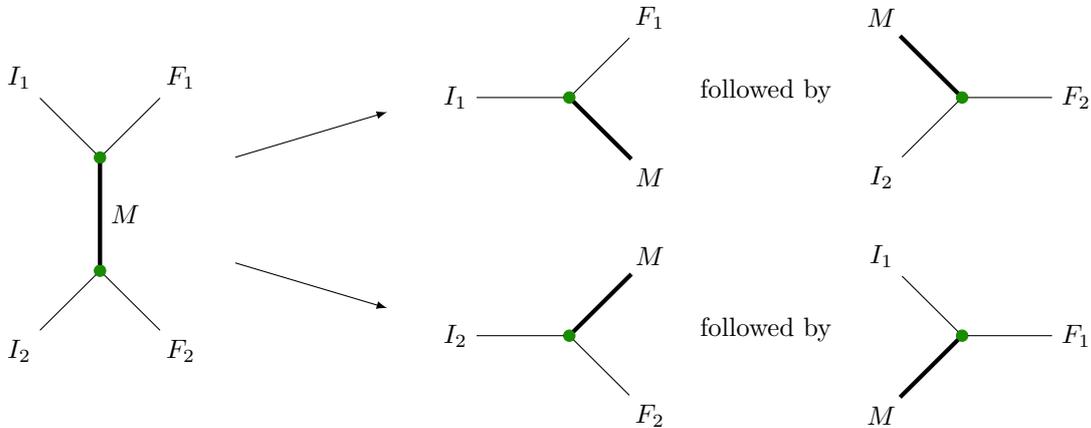
\begin{figure}[H]\begin{center}
	\hfill	
	\raisebox{-0.45\height}{\includegraphics[scale=1]{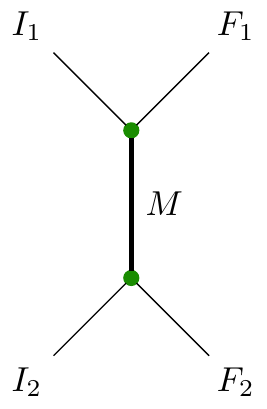}}
	\hfill
	\raisebox{-0.4\height}{
	\begin{tikzpicture}
	\draw[-latex] (0,0.7) -- (2,1.3);
	\draw[-latex] (0,-0.7) -- (2,-1.3);
	\end{tikzpicture}
	}
	\hfill
	\begin{tabular}{c}
	\includegraphics[scale=1]{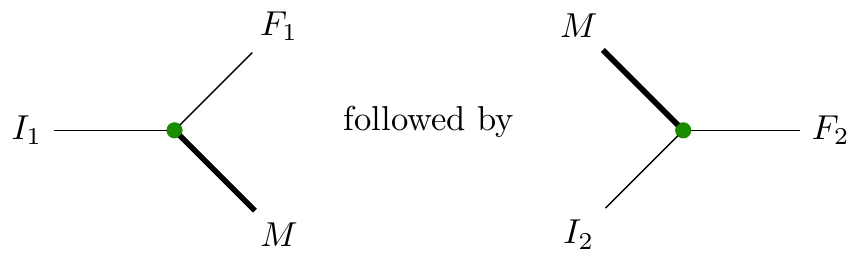}\\[10pt]
	\includegraphics[scale=1]{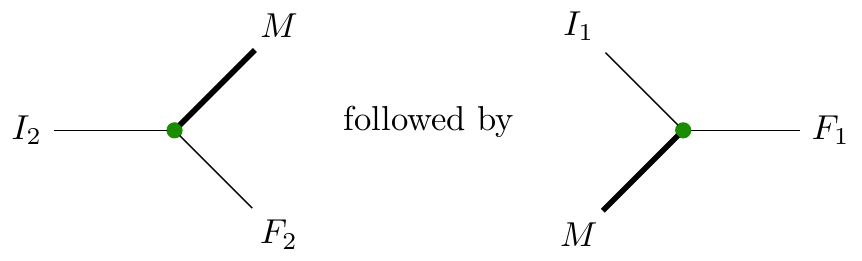}
	\end{tabular}
	\hfill
	\caption{Possible decompositions of a $2\to 2$ $t$-channel process with an on-shell mediator ($M$, distinguished with a thick line), corresponding to \cref{cond:sing}.}
	\label{fig:decomposition}
	\end{center}\end{figure}

In a~recent publication \cite{Grzadkowski:2021kgi} it is shown that for a~$2\to 2$ $t$-channel diagram, for a~given set of particles' masses (two in the initial state, two in the final state, and the mediator's mass), it is possible to determine the range of the CM energy, $\sqrts_1<\sqrts<\sqrts_2$, that leads to appearance of the singularity.
To calculate the thermally averaged cross section, used in cosmological considerations involving Boltzmann equations, one integrates over the CM energy from $\sqrts_{\min}$ (equal to the minimum possible energy of the process, i.e., sum of the initial-state masses or sum of the final-state masses, whichever is greater) to infinity. Hence, the thermally averaged cross section becomes singular whenever the singular range $(\sqrts_1,\sqrts_2)$ intersects with $(\sqrts_{\min},\infty)$, i.e., whenever $\sqrts_{\min}<\sqrts_2$. In the case of a~$2\to 2$ process, it happens when the following conditions are simultaneously satisfied:
	\beg\label[cond]{cond:sing}
	\parbox{0.65\linewidth}
	{
		one of the initial-state particles can decay into\\
		the mediator and the corresponding final-state particle
	}\hspace{-6pt}\\
	\text{and}\\
	\parbox{0.65\linewidth}
	{
		one of the final-state particles can decay into\\
		the mediator and the corresponding initial-state particle,
	}\hspace{-6pt}
	\eeg
where by the ,,corresponding particle'' one should understand the particle connected by the~same vertex to the decaying particle and the~mediator.
In other words, \cref{cond:sing} means that the considered $2\to 2$ process can be decomposed into a sequence of a decay and an inverse decay, as depicted in \cref{fig:decomposition}, with all particles on-shell.

The paper \cite{Grzadkowski:2021kgi} proves the above statement and presents a~method to regularize the $t$-channel singularity. The method bases on applying the real-time (Keldysh-Schwinger) formalism to calculate a~temperature-dependent one-loop mediator's self-energy which is a~result of thermal interactions between the mediator and the surrounding medium. The imaginary part of that self-energy prevents the mediator's propagator from being singular. That method is particularly useful if the singularity appears in cosmological considerations, e.g., those involving hypothetical dark matter particles.

\subsection{This paper}
The aforementioned results of \cite{Grzadkowski:2021kgi} are calculated for the case of a~scalar mediator and scalar loop states. In this work, the method presented there is generalized to the case of particles of spin $0$, $\nicefrac12$, and $1$. The method is illustrated with calculation of particles' self-energies within the Vector-Fermion Dark Matter model \cite{Ahmed:2017dbb}.

The paper is organized as follows.
In \cref{sec:dyson}, a~definition of an~effective width is provided. The effective width is obtained from the imaginary part of the resummed propagator's denominator and expressed in terms of the mediator's self-energy.
\Cref{sec:selfE} contains calculation of the effective width within the Keldysh-Schwinger formalism and presents analytical results depending on particles' spins, expressed using a~model-dependent factor~$X_0$.
In \cref{sec:X0VFDM}, the model-dependent factor $X_0$ is calculated within the Vector-Fermion Dark Matter model \cite{Ahmed:2017dbb}.
\Cref{sec:summary} contains summary and conclusions.
The appendices present Green's functions used in the Keldysh-Schwinger formalism (\cref{app:green}) and contain some details of calculations from the previous sections (\cref{app:krange,app:X0table-VFDM}). \Cref{app:X0table-general} provides the general expression for the aforementioned factor $X_0$, while in \cref{app:weldon} the result obtained in this paper is compared to the one from \cite{Weldon:1983jn}.

\section{Dyson resummation. Effective width in terms of the particle's self-energy}
\label{sec:dyson}
\begin{figure}[H]\begin{center}
\includegraphics[]{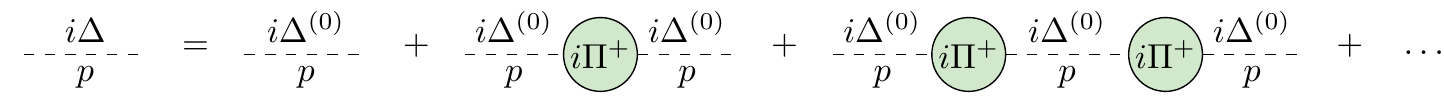}\\
\includegraphics[]{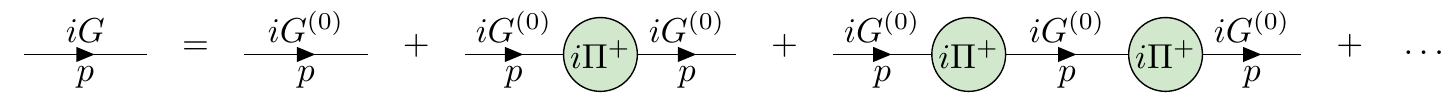}\\
\includegraphics[]{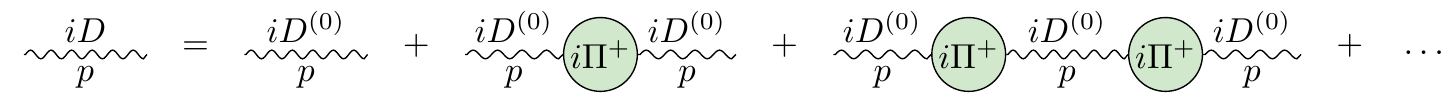}
\caption{Feynman diagrams corresponding to the resummed propagators of a~scalar (top), a~fermion (center), and a~vector (bottom) mediator. The green blobs denote self-energy contributions.}
\label{fig:resummed}
\end{center}\end{figure}

The goal of this section is to calculate a~resummed propagator (see \cref{fig:resummed}) containing an~infinite sum of self-energy corrections. Then, the dependence between the effective width and the self-energy is concluded.

The self-energy $\Pi^+$ used in \cref{sec:selfE} to perform the regularization is the \emph{retarded one-loop} self-energy calculated within the Keldysh-Schwinger formalism, as explained in \cite{Grzadkowski:2021kgi,Czajka:2010zh}. In this section, however, the resummation is performed without any assumptions about the nature of $\Pi^+$.

Let $p$, $T$ denote the momentum of the particle and the temperature of the medium, respectively. For a~scalar (spin-$0$) state, the resummed propagator is given by
	\beq\label{eq:resummationSca}
	i\Delta(p,T)
	&\equiv i\Delta^{(0)}(p)\sum_{n=0}^\infty\Big[i\Pi^+(p,T)i\Delta^{(0)}(p)\Big]^n\\
	&=i\Delta^{(0)}(p)\left[1-i\Pi^+(p,T)i\Delta^{(0)}(p)\right]^{-1}\;,
	\eeq
while for a~fermion (spin-$\nicefrac12$)
	\beq\label{eq:resummationFer}
	iG(p,T)
	&\equiv iG^{(0)}(p)\sum_{n=0}^\infty\Big[i\Pi^+(p,T)iG^{(0)}(p)\Big]^n\\
	&=iG^{(0)}(p)\left[1-i\Pi^+(p,T)iG^{(0)}(p)\right]^{-1}\;,
	\eeq
and for a~vector (spin-$1$)
	\beq\label{eq:resummationVec}
	iD_{\mu\nu}(p,T)
	&\equiv iD_{\mu\alpha}^{(0)}(p)
		\left(\sum_{n=0}^\infty\Big[i\Pi^+(p,T)iD^{(0)}(p)\Big]^n\right)^\alpha_{\,\nu}\\
	&=iD_{\mu\alpha}^{(0)}(p)\,
		\left(\left[1-i\Pi^+(p,T)iD^{(0)}(p)\right]^{-1}\right)^\alpha_{\,\nu}\;.
	\eeq

Here, $\Delta^{(0)},G^{(0)},D^{(0)}$ denote the bare propagator of a~given field (scalar, fermion, or vector, respectively) and $\Pi^+$ is the self-energy (which is calculated in \cref{sec:selfE}). Since particles propagating through a~thermal\footnote{
	\label{ft:T} In fact, the convenient assumption that the medium is in thermal equilibrium is not necessary. The self-energies and, consequently, the effective widths can be calculated for non-thermally distributed energies of the medium's particles as well, assuming different form of distribution functions $n_\text{F,B}$ in \cref{app:green}.
} medium are considered here, the self-energy can depend on the particle's momentum, $p$, and the temperature of the medium, $T$.

Note that for a~scalar or vector state, dimension of the self-energy is energy squared, while for a~fermion state the dimension is the first power of energy.

The following \cref{sec:resummedSca,sec:resummedFer,sec:resummedVec} provide explicit expressions for the resummed propagator in each case. Mass of the particle is denoted by $M$.

\subsection{Scalar state}
\label{sec:resummedSca}
In the scalar case, both $\Delta^{(0)}$ and $\Pi^+$ are scalar quantities:
	\beq\label{eq:treeLevelScaProp}
	\Delta^{(0)}(p)=\frac{1}{p^2-M^2}\;,
	\eeq
so from~\cref{eq:resummationSca} one obtains
	\bbeq
	\Delta(p,T)&=\frac{1}{p^2-M^2+\Pi^+(p,T)}\;.
	\ebeq

\subsubsection{Thermal tree-level propagator vs. the zero-temperature formula}
In general, the tree-level scalar propagator contains a statistical part proportional to ${\delta(p^2-m^2)}$ (see, e.g., appendix~A.4 of \cite{Bellac:2011kqa}):
	\beq
	\Delta^{(0)}_\text{th}(p)
	&\equiv \frac{1}{p^2 - M^2} - 2\,\pi\,i\,n(E_p)\,\delta(p^2-M^2)\\
	&= \Delta^{(0)}(p)\left[1 - 2\,\pi\,i\,n(E_p)\,(p^2-M^2)\,\delta(p^2-M^2)\right]\;,
	\eeq
where $n(E_p)$ is the distribution function:
	\beq
	n(E_p) \equiv [e^{\beta E_p} - 1]^{-1}\;,\qquad E_p\equiv\sqrt{\vec p^2+m^2}\;,
	\eeq
and $\Delta^{(0)}(p)$ is the propagator given by \cref{eq:treeLevelScaProp}. Then, the resummed propagator calculated using the $\Delta_\text{th}^{(0)}$ bare propagator becomes:
	\beq
	\Delta
	&= \Delta_\text{th}^{(0)}\left[1 + \Pi^+ \Delta_\text{th}^{(0)}\right]^{-1}\\
	&= \Delta^{(0)}\,
	\left[1 - 2\,\pi\,i\,n(E_p)\,(p^2-M^2)\,\delta(p^2-M^2)\right]\\
	&\mkern40mu\times
	\left[1 + \Pi^+ \Delta^{(0)} - 2\,\pi\,i\,\Pi^+\,n(E_p)\,\delta(p^2-M^2)\right]^{-1}\\
	&= \frac{1 - 2\,\pi\,i\,n(E_p)\,(p^2-M^2)\,\delta(p^2-M^2)}
		{p^2-M^2 + \Pi^+ - 2\,\pi\,i\,\Pi^+\,n(E_p)\,(p^2-M^2)\,\delta(p^2-M^2)}\;.
	\eeq
As long as $\Pi^+$ is regular (i.e., finite, non-zero and smooth enough) at $p^2=M^2$, all the above operations are legal and the {$(p^2-M^2)\,\delta(p^2-M^2)$} component can be dropped\footnote{
This is because $f(x)=x\,\delta(x)$ is equivalent to zero in the distributional sense. Hence, as long as it is multiplied by regular functions only, such a term must vanish in comparison to any non-zero term.
} both in the numerator and the denominator of the above expression, which provides the same result as obtained for the zero-temperature bare propagator \labelcref{eq:treeLevelScaProp}. Therefore, the statistical component could have been neglected from the very beginning. The same logic applies to the fermion and vector cases.

\subsection{Fermion state}
\label{sec:resummedFer}
In the fermion case, $G^{(0)}$ and $\Pi^+$ posses spinor structure. They can\footnote{For a~brief discussion, see section~A of chapter~III in \cite{Das:2017vfh}.} be expressed using the gamma matrices as
	\beq
	G^{(0)}(p)&=\frac{1}{\sl p-M}\;,\\
	\Pi^+(p,T)&=\left[A_v(p,T)+A_a(p,T)\gamma_5\right]\,\sl p
		+\left[B_v(p,T)+B_a(p,T)\gamma_5\right]\,M\;,
	\eeq
where $A_{v,a}$ and $B_{v,a}$ are dimensionless scalar quantities and $\sl p\equiv p^\mu\gamma_\mu$. According to~\cref{eq:resummationFer}, the resummed propagator is given by
	\beq
	iG
	&=iG^{(0)}\left[1-i\Pi^+\,iG^{(0)}\right]^{-1}\\
	&=i\,\frac{1}{\sl p-M}
		\left[\frac{(1+A_v+A_a\,\gamma_5)\,\sl p+(-1+B_v+B_a\,\gamma_5)\,M}{\sl p-M}\right]^{-1}\\
	&=i\,\frac{(1+A_v+A_a\,\gamma_5)\,\sl p+(1-B_v+B_a\,\gamma_5)\,M}
		{\left[(1+A_v)^2-A_a^2\right]p^2-\left[(1-B_v)^2-B_a^2\right]M^2}\;.
	\eeq
Hence,
	\beq
	G(p,T)
	&=\frac{[1+A_v(p,T)+A_a(p,T)\,\gamma_5]\,\sl{p}+[1-B_v(p,T)+B_a(p,T)\,\gamma_5]\,M}
	{\left(\left[1+A_v(p,T)\right]^2-A_a(p,T)^2\right)p^2
		-\left(\left[1-B_v(p,T)\right]^2-B_a(p,T)^2\right)M^2}\;.
	\eeq
To simplify this formula, let us assume that:
	\begin{itemize}
	\item the self-energy is small comparing to the mass: $|A_{v,a}|\ll1$, $|B_{v,a}|\ll 1$,
	\item the axial coefficients of the self-energy are at most of the order of the vector coefficients: $|A_a|,|B_a|\lesssim|A_v|,|B_v|$,
	\item the singular propagator is almost on-shell:\footnote{This is assumed basing on the fact that the vicinity of the singular point dominates the momentum-space integral.} $|p^2-M^2|\ll M^2$.
	\end{itemize}
Under these assumptions one obtains\footnote{This result agrees with eq.~(11) of \cite{Gonchar:2006xv}. Note that their $A$, $B$ are here $-B_v M$, $-A_v$, respectively.}
	\beq
	G(p,T)
	&\simeq\frac{\sl{p}+M}{p^2-M^2+2\left[A_v(p,T)+B_v(p,T)\right]\,M^2}\;.
	\eeq
Note that $A_v$ and $B_v$ can be calculated from the self-energy using the trace operator:
	\beq
	A_v(p,T)&=\frac{1}{4p^2}\tr[\sl{p}\Pi^+(p,T)]\;,&
	B_v(p,T)&=\frac{1}{4M}\tr[\Pi^+(p,T)]\;,
	\eeq
so
	\bbeq
	G(p,T)&\simeq\frac{\sl{p}+M}{p^2-M^2+2\tr\!\left[\frac{\sl p+M}{4}\Pi^+(p,T)\right]}\;.
	\ebeq

\subsection{Vector state}
\label{sec:resummedVec}
In the vector case, $D^{(0)}$ and $\Pi^+$ have Lorentz structure and can be conveniently expressed as
	\beq
	D_{\mu\nu}^{(0)}(p)
	&=-\left(\frac{T_{\mu\alpha}}{p^2-M^2}-\frac{L_{\mu\alpha}}{M^2}\right)\;,\\
	\Pi^+_{\mu\nu}(p)
	&=\Pi_\text{T}(p,T)\,T_{\mu\nu}
		+\Pi_\text{L}(p,T)\,L_{\mu\nu}\;,
	\eeq
where $\Pi_\text{T}$ and $\Pi_\text{L}$, denoting the transverse and the longitudinal component of the self-energy, respectively, are scalar quantities of dimension of energy squared. The transverse projector $T_{\mu\nu}$ and the longitudinal projector $L_{\mu\nu}$ are defined as
	\beg
	T_{\mu\nu}\equiv g_{\mu\nu}-\frac{p_\mu p_\nu}{p^2}\;,
	\mkern80mu
	L_{\mu\nu}\equiv \frac{p_\mu p_\nu}{p^2}\;.
	\eeg
Then, according to~\cref{eq:resummationVec},
	\beq
	iD_{\mu\nu}
	&=iD_{\mu\alpha}^{(0)}
		\left(\left[1-i\Pi^+(p,T)iD^{(0)}(p)\right]^{-1}\right)^\alpha_{\,\nu}\\
	&=-i\,\left(\frac{T_{\mu\alpha}}{p^2-M^2}-\frac{L_{\mu\alpha}}{M^2}\right)
		\left(\left[\frac{p^2-M^2-\Pi_T}{p^2-M^2}T
		+\frac{M^2+\Pi_L}{M^2}L
		\right]^{-1}\right)^\alpha_{\,\nu}\\
	&=-i\,\left(\frac{T_{\mu\alpha}}{p^2-M^2}-\frac{L_{\mu\alpha}}{M^2}\right)
		\left[\frac{p^2-M^2}{p^2-M^2-\Pi_T}T^\alpha_\nu
		+\frac{M^2}{M^2+\Pi_L}L^\alpha_\nu
		\right]\\
	&=-i\,\frac{T_{\mu\nu}}{p^2-M^2-\Pi_T}+i\frac{L_{\mu\nu}}{M^2+\Pi_L}\\
	&=i\,\frac{-g_{\mu\nu}+\frac{p_\mu p_\nu}{M^2+\Pi_L}\frac{p^2-\Pi_T+\Pi_L}{p^2}}
		{p^2-M^2-\Pi_T}\;,
	\eeq
so\footnote{This result agrees with eq.~(20) of \cite{Nowakowski:1993iu}, up to the~sign of $\Pi_T$ in their equation. Note that their $\Pi_T(q^2)$ is equal to $\Pi_T(q^2)$ used here, while their $\Pi_L(q^2)$ is here $[\Pi_L(q^2)-\Pi_T(q^2)]/q^2$.}
	\beq
	D_{\mu\nu}(p,T)
	&=\frac{-g_{\mu\nu}
		+\frac{p_\mu p_\nu}{M^2+\Pi_L(p,T)}\frac{p^2-\Pi_T(p,T)+\Pi_L(p,T)}{p^2}
		}{p^2-M^2-\Pi_T(p,T)}\;.
	\eeq
Assuming that the self-energy is small (i.e., $|\Pi_L|,|\Pi_T|\ll M^2$) and the propagating state is almost on-shell (so that ${|p^2-M^2|\ll M^2}$), one obtains
	\beq
	D_{\mu\nu}(p,T)\simeq\frac{-g_{\mu\nu}+\frac{p_\mu p_\nu}{M^2}}{p^2-M^2-\Pi_T(p,T)}\;.
	\eeq
Quantity $\Pi_T$ can be expressed as
	\beq
	\Pi_T(p,T)=\frac13\left(g^{\mu\nu}-\frac{p^\mu p^\nu}{p^2}\right)\Pi^+_{\mu\nu}(p,T)\;,
	\eeq
so
	\bbeq
	D_{\mu\nu}(p,T)\simeq\frac{-g_{\mu\nu}+\frac{p_\mu p_\nu}{M^2}}{p^2-M^2
		+\frac13\left(-g^{\alpha\beta}+\frac{p^\alpha p^\beta}{p^2}\right)\Pi^+_{\alpha\beta}(p,T)}\;.
	\ebeq

\subsection{Effective width}
The imaginary part of the expression that is present in the resummed propagator but absent in the~free propagator can be denoted as $\Sigma(p,T)$:
	\beq\label{eq:sigma}
	\Sigma(p,T)\equiv
	\begin{cases}
		\im\Pi^+(p,T)
			&\text{ scalar case}\\
		\im\left(
			\tr\!\left[\frac{\sl p+M}{2}\Pi^+(p,T)\right]
			\right)
			&\text{ fermion case}\\
		\im\left[\frac13\left(-g^{\mu\nu}+\frac{p^\mu p^\nu}{p^2}\right)\Pi^+_{\mu\nu}(p,T)\right]
			&\text{ vector case}
	\end{cases}\;.
	\eeq
In each case, in an~analogy to the Breit-Wigner propagator, a~$(p,T)$-dependent effective decay width $\Gamma_\eff(p,T)$ can be introduced in the following way:
	\beq\label{eq:geff}
	\Gamma_\eff(p,T)\equiv\frac{|\Sigma(p,T)|}{M}\;.
	\eeq
If this quantity is non-zero, it regularizes the singular on-shell propagator in the very same manner as the Breit-Wigner propagator is regularized by the standard decay width.

Note that the real part of the expression changes the value of mass, e.g., the bare mass $M^2$ becomes {$M^2 - \Re\Pi^+(p,T)$} in the scalar case. Hence, strictly speaking, from now on, $M^2$ denotes the dressed mass. Nevertheless, it is assumed that the shift is small in comparison to the bare mass $M^2$, so that the kinematics of the process involving the considered particle is not affected qualitatively (in particular, \cref{cond:sing} still applies if did before).

\section{Calculation of the regulator \topstr{$\Sigma(p,T)$}{Sigma(p,T)}}
\label{sec:selfE}

\begin{figure}[H]\begin{center}
	\raisebox{-0.55\height}{\includegraphics[scale=1]{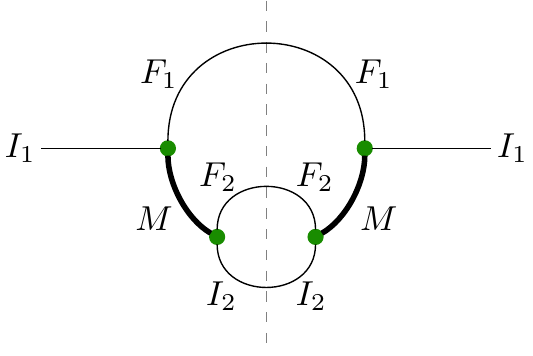}}
	\raisebox{-0.5\height}{
	\hspace{1cm}
	\begin{tikzpicture}
	\draw[-latex] (0,0) -- (2,0);
	\draw (3.5,-1.5) -- (3.5,1.5);
	\end{tikzpicture}
	}
	\raisebox{-0.5\height}{\includegraphics[scale=0.75]{./diagrams/diag.pdf}}
	\raisebox{-0.5\height}{
	\begin{tikzpicture}
	\draw (0,-1.5) -- (0,1.5);
	\node at (0.3,1.5) {2};
	\end{tikzpicture}
	}
	\caption{Relation between the self-energy of particle $I_1$ and the amplitude of the considered $t$-channel process $I_1 I_2 \to F_1 F_2$. The dashed line represents the corresponding cut of the self-energy diagram. For clarity, the line representing the $t$-channel mediator is thickened.}
	\label{fig:KB-selfE}
	\end{center}\end{figure}

In the statistical field theory, the Boltzmann equation appears as a semi-classical approximation of the so-called Kadanoff-Baym equations \cite{Kadanoff2018} (for a derivation of the Boltzmann equation see also section~10 of \cite{Mrowczynski:1989bu}), which are equations of motion of thermal Green's functions. The amplitude of the discussed $t$-channel process shown in \cref{fig:decomposition} enters those equations as a part of the contribution corresponding to the self-energy of one of the external particles. \Cref{fig:KB-selfE} shows the relation between the self-energy of particle $I_1$ and the amplitude of the process. The propagator of the mediator, singular in the usual treatment, has to be replaced by the statistical counterpart. In this paper, this is achieved by including in the resummed propagator the one-loop retarded self-energy calculated within the Keldysh-Schwinger formalism.
This section provides a~result for the effective width $\Gamma_\eff\equiv|\Sigma|/M$, defined by \cref{eq:geff}, obtained for $\Pi^+$ denoting the thermal self-energy.

\begin{figure}[H]\begin{center}
	\includegraphics[width=4cm]{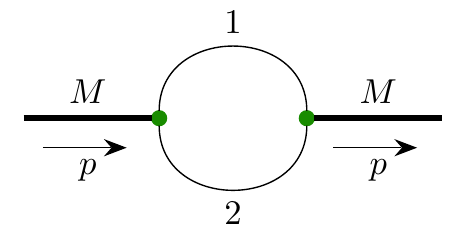}
	\caption{Diagram of the mediator's one-loop self-energy.}
	\label{fig:loop}
	\end{center}\end{figure}

Let us denote the mediator's mass by $M$, its four-momentum measured in the rest frame of the medium by $p=(p_0,\vec p)$, and the states present in the loop by~$1$ and~$2$ (with masses~$m_1$ and~$m_2$, respectively), see \cref{fig:loop}.

For convenience, one can define
	\beq
	\beta &\equiv \frac{1}{T}\;,&
	E_p&\equiv\sqrt{\vec p^2+M^2}\;,&
	E_{1,2}&\equiv\sqrt{\vec k^2+m_{1,2}^2}\;.
	\eeq	
Because the process is expected to be $t$-channel singular, it is assumed that the mediator is (almost) on-shell, so
	\beg
	p_0 = E_p\;,\qquad
	p^2=M^2\;,
	\eeg
and stable in vacuum:
	\beq\label[ineq]{ineq:stable}
	M<m_1+m_2\;.
	\eeq
If all three states are scalars and the vertex factor is $\mu$, the mediator's retarded one-loop self-energy $\Pi^+ (x)$ calculated within the Keldysh-Schwinger formalism can be found (see \cite{Grzadkowski:2021kgi}) as
	\beq\label{eq:selfE}
	\Pi^+(p,T) = \frac{i}{2} 
	 \int \frac{d^4k}{(2\pi)^4} \Big[\mu\,\Delta_1^+(k+p) \,\mu\, \Delta_2^\sym (k,T)
	+ \mu\,\Delta_1^\sym(k,T) \,\mu\, \Delta_2^-(k-p) \Big]\;.
	\eeq

If the particles (the mediator and the loop states) have non-zero spins, the self-energy can be found analogously, using $G$ ($D_{\mu\nu}$) as a~fermion (vector) propagator and replacing $\mu$ by an~appropriate vertex contribution.

The retarded, advanced, and symmetric Green's functions for scalar, fermion, and vector particles: $\Delta^\pm (p)$, $G^\pm (p)$, $D_{\mu\nu}^\pm (p)$, and $\Delta^\sym(p,T)$, $G^\sym(p,T)$, $D_{\mu\nu}^\sym(p,T)$ are provided in \cref{app:green}. Substituting them into \cref{eq:selfE} will allow us to obtain $\Pi^+(p,T)$ needed to calculate $\Sigma$ defined by \cref{eq:sigma}.

According to~\cref{eq:sigma}, in the case of a~scalar mediator the result of \cref{eq:selfE} is what is needed to calculate $\Sigma$. In the case of a~fermion mediator one has to multiply the result by $\frac{\sl p+M}{2}$ and calculate the trace, while in the case of a~vector mediator the result has to be multiplied by $\frac13\left(-g^{\mu\nu}+\frac{p^\mu p^\nu}{p^2}\right)$. Then, in each case, the imaginary part should be found. Regardless of the case, the effect of these manipulations can be expressed as
	\beq\label{eq:here}
	\Sigma(p,T)
		= \im\Bigg[
		\frac{1}{2}
		\int \frac{d^4k}{(2\pi)^4}
		\bigg\{
			&\phantom{+}\frac{X_{(p^2,\,k^2,\,(k+p)^2)}}{(k+p)^2 - m_1^2 + i\, {\rm sgn}(k_0+p_0)\,0^+}\\
			&\times\frac{\pi}{E_2}\Big( \delta (E_2 - k_0) 
			+ \delta (E_2 + k_0)\Big)\,f_2(\beta E_2)\\
			&+\frac{X_{(p^2,\,(k-p)^2,\,p^2)}}{(k-p)^2 - m_2^2 - i\, {\rm sgn}(k_0-p_0)\,0^+}\\
			&\times\frac{\pi}{E_1}\Big( \delta (E_1 - k_0) 
			+ \delta (E_1 + k_0)\Big)\,f_1(\beta E_1)
		\;\bigg\}
		\Bigg]\;,
	\eeq
where function $f_i$ is defined as
	\beq
	f_i(x)
	&\equiv
		\begin{cases}
			\frac{e^x-1}{e^x+1}	&\text{ if particle $i$ is a~fermion}\\
			\frac{e^x+1}{e^x-1}	&\text{ if particle $i$ is a~boson}
		\end{cases}\;,
	\eeq
while the Lorentz invariant $X$, being a~product of appropriate coupling constants and numerators of the propagators after applying the procedure mentioned above \cref{eq:here}, has to be calculated within a~given model.

The integration over $d^4k$ makes the result insensitive to the direction of $\vec p$, so the imaginary part of the self-energy becomes energy-dependent:
	\beq
	\Sigma(p,T)=\Sigma(E_p,T)\;.
	\eeq
Integrating over $k_0$ one finds
	\beq
	\Sigma(E_p,T)
	= \im\Bigg[\;
	\frac{1}{4}
	\int \frac{d^3k}{(2\pi)^3}
	\Bigg\{
		\phantom{+}\frac{X_{(p^2,\,k^2,\,(k+p)^2)}}{E_2}\, f_2(\beta E_2)
		\mkern-115mu &\\
		\times\bigg[
			\phantom{+}&\frac{1}{M^2+m_2^2-m_1^2+2pk + i\, {\rm sgn}(E_2+E_p)\,0^+}\\
			+&\,\frac{1}{M^2+m_2^2-m_1^2+2pk + i\, {\rm sgn}(-E_2+E_p)\,0^+}
		\,\bigg]\\
		+\,\frac{X_{(p^2,\,(k-p)^2,\,k^2)}}{E_1}\,f_1(\beta E_1)
		\mkern-115mu &\\
		\times\bigg[
			\phantom{+}&\frac{1}{M^2-m_2^2+m_1^2-2pk - i\, {\rm sgn}(E_1-E_p)\,0^+}\\
			+&\,\frac{1}{M^2-m_2^2+m_1^2-2pk - i\, {\rm sgn}(-E_1-E_p)\,0^+}
		\,\bigg]
	\;\Bigg\}
	\Bigg]\;.
	\eeq
Due to the Sochocki relation:
	\beq
	\lim_{\eps\to 0^+}\frac{1}{x \pm i\eps} = {\cal P}\frac{1}{x } \mp i \pi \delta(x),
	\eeq
the result equals to
	\beq
	\Sigma(E_p,T)
		= \frac{\pi}{4}
		\int \frac{d^3k}{(2\pi)^3}
		\Bigg\{
		-	X_0\, f_2(\beta E_2)
			&\,\bigg[\,\frac{{\rm sgn}(E_2+E_p)}{E_2} + \frac{{\rm sgn}(-E_2+E_p)}{E_2}\,\bigg]\\
			&\times\delta(M^2+m_2^2-m_1^2+2pk)\\
			X_0\, f_1(\beta E_1)
			&\,\bigg[\,\frac{{\rm sgn}(E_1-E_p)}{E_1} + \frac{{\rm sgn}(-E_1-E_p)}{E_1}\,\bigg]\\
			&\times\delta(M^2-m_2^2+m_1^2-2pk)
		 \Bigg\},
	\eeq
where $X_0$ is the on-shell value of $X$:
	\beq\label{eq:XY}
	X_0	&\equiv	X_{(M^2,\,m_2^2,\,m_1^2)}\;.
	\eeq

For further calculations, the spherical coordinates with the $z$ axis set along the vector $\vec p$ are used: instead of $d^3k$, integration over $\vec k^2\,d\n k\,d\cos\theta\,d\phi$ is performed, with the angles $\theta$ and $\phi$ defined via the following relations:
	\beq
	p^\mu&=(E_p=\sqrt{\vec p^2+M^2},\,0,\,0,\,\n p)\;,\\
	k^\mu&=(k_0,\,\n k\sin\theta\cos\phi,\,\n k\sin\theta\sin\phi,\,\n k\cos\theta)\;.
	\eeq
In this coordinate system, the Lorentz invariants are given by
	\beq
	p^\mu p_\mu	&= M^2\;,&
	k^\mu k_\mu	&= k_0^2-\vec k^2\;,&
	p^\mu k_\mu &= E_pk_0-\n k\n p\cos\theta\;.
	\eeq

It is now assumed that $\vec p\neq 0$.\footnote{The other case is easy to solve and appears to be a~limit of this one; the result is provided in \cref{eq:selfEFinal-p0}.} The integral over the azimuthal angle~$\phi$ is trivial and the remaining integrals are
	\beq\label{eq:selfE1}
	\Sigma(E_p,T)
	&=\begin{aligned}[t]
	\frac{X_0}{32\pi} 
		\int_0^\infty \vec k^2 d\n k\int_{-1}^1 d\,\cos\theta\;
	\mkern-100mu &\\
	\times\Bigg\{
	-	f_2(\beta E_2)\,
		\bigg[
	&\phantom{+}\,
			\frac{{\rm sgn}\big(E_2 + E_p\big)}{E_2}\,
			\frac{\delta (\cos\theta - \cos\alpha_2)}{\n k\n p}\\
	&		+\frac{{\rm sgn}\big(-E_2 + E_p\big)}{E_2}\,
			\frac{\delta (\cos\theta - \cos\beta_2)}{\n k\n p}
		\,\bigg]\\
	+	f_1(\beta E_1)\,
		\bigg[
	&\phantom{+}\,
			\frac{{\rm sgn}\big(E_1 - E_p\big)}{E_1} \,
			\frac{\delta (\cos\theta - \cos\alpha_1)}{\n k\n p}\\
	&		+\frac{{\rm sgn}\big(-E_1 - E_p\big)}{E_1}\,
			\frac{\delta (\cos\theta - \cos\beta_1)}{\n k\n p}
		\,\bigg]
	\Bigg\}
	\end{aligned}\\
	&=\begin{aligned}[t]
	\frac{X_0}{32\pi\n p}
	\bigg\{
	-	 \int_0^\infty dE_2 \int_{-1}^1 d\,\cos\theta\,
		f_2(\beta E_2)\mkern-100mu &\\
	\times\big[
			\phantom{+}\,
		&	\sgn(E_2 + E_p)\,\delta (\cos\theta - \cos\alpha_2)\\
		+\,&	\sgn(-E_2 + E_p)\,\delta (\cos\theta - \cos\beta_2)
		\big]\\
	+	 \int_0^\infty dE_1 \int_{-1}^1 d\,\cos\theta\,
		f_1(\beta E_1)\mkern-100mu &\\
		\times\big[
			\phantom{+}\,
		&	\sgn(E_1 - E_p)\,\delta (\cos\theta - \cos\alpha_1)\\
		+\,&\sgn(-E_1 - E_p)\,\delta (\cos\theta - \cos\beta_1)
		\big]
	\bigg\}
	\end{aligned}
	\eeq
where
	\beq\label{eq:cos}
	\cos\alpha_1
		&\equiv \frac{-(m_1^2 - m_2^2 + M^2) + 2E_1E_p}{2\,\n k\n p}\;,&
	\cos\beta_1
		&\equiv \frac{-(m_1^2 - m_2^2 + M^2) - 2E_1E_p}{2\,\n k\n p}\;,\\
	\cos\alpha_2
		&\equiv \frac{-(m_1^2 - m_2^2 - M^2) + 2E_2E_p}{2\,\n k\n p}\;,&
	\cos\beta_2
		&\equiv \frac{-(m_1^2 - m_2^2 - M^2) - 2E_2E_p}{2\,\n k\n p}\;.
	\eeq

Since the integrand depends on $\cos\theta$ via $\delta(\cos\theta-\cos\alpha_i)$ or $\delta(\cos\theta-\cos\beta_i)$ functions ($i=1,2$) only, the integration over $d\cos\theta$ effectively limits the range of $E_{1,2}$ to the values leading to $|\cos\alpha_i|<1$ or $|\cos\beta_i|<1$. In \cref{app:krange}, it is proved that this range is non-empty only if
	\beq\label[ineq]{ineq:massCond}
	m_1>m_2+M
	\eeq
or
	\beq\label[ineq]{ineq:massCondAlternative}
	m_2>m_1+M\;.
	\eeq
Because the existence of a~particle that is allowed to decay into the mediator and another particle is necessary for the singularity to occur (see \cref{cond:sing}), it is possible to find two states of masses $m_1$ and $m_2$ that satisfy one of the above conditions. Hence, if the $t$-channel singularity occurs, it can always be regularized using the method presented in this paper. Without loss of generality, one can assume that \cref{ineq:massCond} holds, so the particle of mass $m_1$ can decay as $m_1 \rightarrow m_2 + M$. Also the inverse process $m_2 + M \rightarrow m_1$ is kinematically allowed. Note that in \cref{sec:discussion} it is shown that in the limit of $m_1=m_2+M$ the thermal self-energy vanishes.

As shown in \cref{app:krange}, given that \cref{ineq:massCond} holds, the terms containing deltas with $\cos\beta_{1,2}$ vanish and the remaining part can be expressed as
	\beq\label{eq:deltacos}
	\Sigma(E_p,T)
	= \frac{X_0 }{32\pi}\frac{1}{\n p}
		\bigg[
		&-\int_{b-a-E_p}^{b+a-E_p}\,dE_2\,f_2(\beta E_2)\,\sgn(E_2 + E_p)\\
		&+\int_{b-a}^{b+a}\,dE_1\,f_1(\beta E_1)\,\sgn(E_1 - E_p)
		\;\bigg]\;,
	\eeq
with $\ca$ and $\cb$ defined as
	\beg
	\ca\equiv\frac{\lambda(m_1^2,m_2^2,M^2)^{1/2}}{2M^2}\,\n p\;,\qquad
	\cb\equiv\frac{m_1^2-m_2^2+M^2}{2M^2}\,E_p\;,\\
	\lambda(m_1^2,m_2^2,M^2)\equiv\left[m_1^2-(m_2-M)^2\right]\left[m_1^2-(m_2+M)^2\right]\;.
	\eeg
The sgn function in the first integral of \cref{eq:deltacos} gives, obviously, $+1$, and since
	\beq	\label[ineq]{ineq:abe}
	\cb -\ca -E_p>0\;,
	\eeq
(see~\cref{eq:positive}), the second sgn function gives $+1$ as well. Therefore, the imaginary part of the self-energy becomes
	\beq\label{eq:selfE-almostFinal}
	\Sigma(E_p,T)
	&= \frac{X_0 }{32\pi}\frac{1}{\n p}
		\bigg[\;
		-\int_{b-a-E_p}^{b+a-E_p}\,dE_2\,f_2(\beta E_2)
		+\int_{b-a}^{b+a}\,dE_1\,f_1(\beta E_1)
		\;\bigg]\\
	&= \frac{X_0 }{32\pi}\frac{1}{\beta\n p}
	\bigg[\;
	\int_{\beta(\cb -\ca)}^{\beta(\cb +\ca)}
		f_1(x)\,dx
	-\int_{\beta(\cb -\ca -E_p)}^{\beta(\cb +\ca -E_p)}
		f_2(x)\,dx
	\;\bigg]\;,
	\eeq
where
	\beq
	f_i(x) \equiv
	\begin{cases}
	\frac{e^x+1}{e^x-1}&\text{ particle $i$ is a~boson}\\
	\frac{e^x-1}{e^x+1}&\text{ particle $i$ is a~fermion}
	\end{cases}\;,\qquad i=1,2\;.
	\eeq
After integration, the final result, depending on the spins of the particles in the loop, is given as

	\bbeq
	\Gamma_\eff(E_p,T)
	&\equiv\frac{1}{M}|\Sigma(E_p,T)|\;,\\
	\Sigma(E_p,T)
	&=	\frac{1}{16\pi}\;\frac{X_0}{\beta\n p}\;
		\Bigg[
			\ln\frac{ e^{\beta(\cb +\ca)}				+\eta_1 }{ e^{\beta(\cb -\ca)}				+\eta_1 }
		-	\ln\frac{ e^{\beta(\cb +\ca)}e^{-\beta E_p}	+\eta_2 }{ e^{\beta(\cb -\ca)}e^{-\beta E_p}	+\eta_2 }
		\Bigg]\\
	&=	\frac{1}{16\pi}\;\frac{X_0}{\beta\n p}\;
		\ln\Bigg[
			1 + \frac
			{
				e^{-\beta(\cb -\ca)} e^{\beta E_p}
				\left(1 - e^{-2\beta a}\right)
				\left(\eta_2 - \eta_1\,e^{-\beta E_p}\right)
			}
			{
				\left(1 + \eta_1\,e^{-\beta(\cb -\ca)}\right)
				\left(1 + \eta_2\,e^{-\beta(\cb +\ca)}e^{\beta E_p}\right)
			}
		\Bigg]\;,
	\label{eq:selfEFinal}
	\ebeq
where
	\beg
	\eta_i\equiv
	\begin{cases}
	-1&\text{particle $i$ is a~boson}\\
	+1&\text{particle $i$ is a~fermion}
	\end{cases}\;,\\
	\beta\equiv \frac{1}{T}\;,\qquad
	|\vec p|\equiv\sqrt{E_p^2-M^2}\;,\\
	\ca\equiv\frac{\n p\,\sqrt{\lambda(m_1^2,m_2^2,M^2)}}{2M^2}\;,\qquad
	\cb\equiv\frac{m_1^2-m_2^2+M^2}{2M^2}\,E_p\;,\\
	\lambda(m_1^2,m_2^2,M^2)\equiv\left[m_1^2-(m_2-M)^2\right]\left[m_1^2-(m_2+M)^2\right]\;,
	\eeg
$M$ denotes the mediator's mass and $X_0$ is defined in \cref{eq:XY}. Values of $X_0$ are provided, for the general case, in \cref{app:X0table-general}, and for the Vector-Fermion Dark Matter model \cite{Ahmed:2017dbb} in \cref{app:X0table-VFDM,sec:X0VFDM}.

\subsection{Result discussion}
\label{sec:discussion}
The above result is consistent with that derived in \cite{Weldon:1983jn} up to a~spin-dependent factor of $\nicefrac12$ for a~fermion mediator and $\nicefrac13$ for a~vector mediator, see \cref{app:weldon}.

As it is clear from the last line of \cref{eq:selfEFinal}, the logarithmic part is positive if the particle ,,2'' is a~fermion and negative otherwise, given that $m_1>m_2+M$ (\cref{ineq:massCond}) and $b>a+E_p$ (\cref{ineq:abe}). This corresponds to the sign of the factor $X_0$ calculated in \cref{app:X0table-general}. Consequently, in any model consisting of scalars, fermions and vectors, the regulator $\Sigma$ is always positive and regularizes the singularity (therefore, the absolute value in \cref{eq:selfEFinal} can be omitted).

If $m_1=m_2+M$, quantity $\ca$ vanishes and so does the effective width (as expected, since non-zero effective width is a~consequence of a~decay of particle ,,1'' into ,,2'' and the mediator, see \cref{ineq:massCond}).

Note that values of \cref{eq:selfEFinal} for $\eta_1=-\eta_2=1$ and for $\eta_1=-\eta_2=-1$ are not identical, even though both assume a~boson and a~fermion in the loop. The difference is that in the first case the particle allowed to decay is the~fermion, while in the second case the decaying particle is the~boson, which leads to a~different statistical factor that has to be taken into account in each case.

In the limit of $\beta\to\infty$ (zero-temperature limit), the result tends to~$0$ (the argument of the logarithmic function in the last line becomes $1$). This behaviour is expected, as the thermal width is a~result of interactions between the mediator and a~medium of non-zero temperature. Taking $\beta\to\infty$ reflects lack of the medium.

The same happens in the limit of $\n p\to\infty$. Physically, the process is regularized by interactions between the mediator and the thermal bath. The calculated self-energy represent destruction of the mediator due to interaction with particle ,,2'' (from the thermal bath), with production of particle ,,1'', and a subsequent decay of particle ,,1'' into particle ,,2'' and mediator of exactly the same energy as the destroyed one.
	\Cref{eq:e2} provides the minimal value of energy $E_2$ necessary for an on-shell production of particle ,,$1$'' in a $M,2\to 1$ process (with the mediator having energy $E_p$), which is $b-a-E_p$. As $\n p$ goes to infinity, this minimal energy tends to infinity as well, so only a small amount of particles in the thermal bath is energetic enough. This leads to statistical suppression of the regulator.

In the limit of $\vec p\to0$ (zero-momentum limit), factor $\ca$ is equal to $0$, so the logarithmic function present in the result also tends to $0$. However, $p$ appears not only in~$\ca$ but also divides the results, so the overall limit can be non-trivial. In fact, it is equal to:
	\beq
	\label{eq:selfEFinal-p0}
	\Gamma_\eff(\vec p=0,T)
	&=	\frac{X_0}{16\pi\,M}\;
		\frac{ \lambda(m_1^2,m_2^2,M^2)^{1/2} }{M^2}\;
		\frac{
			e^{-\beta(b_0 - M)}\;\left( \eta_2 - \eta_1\,e^{-\beta M} \right)
		}{
			\left(	1 + \eta_1\,e^{-\beta b_0}	\right)
			\left(	1 + \eta_2\,e^{-\beta(b_0 - M)}	\right)
		}\;,\\
	\eeq
with $b_0 \equiv (m_1^2-m_2^2+M^2)/(2M) > M$.

If the temperature is small enough to ensure $\beta(\cb-\ca-E_p)\gtrsim 3$ (so that $e^{\beta(\cb-\ca-E_p)}\gg 1$), \cref{eq:selfEFinal} can be expanded around $e^{-\beta(\cb-\ca-E_p)}=0$, giving
	\beq
	\Gamma_\eff(E_p,T)
	&\simeq
		\frac{1}{16\pi\,M}\;\frac{X_0}{\beta\n p}\;
		e^{-\beta(\cb -\ca -E_p)}\left(1-e^{-2\,\beta\ca}\right)\left(\eta_2 -\eta_1\,e^{-\beta E_p}\right)\;.\\
	\eeq
This form is especially useful for numerical calculations.

\section{Values of the factor \topstr{$X_0$}{X0} calculated within the VFDM model}
\label{sec:X0VFDM}

In this section, the presented regularization method is illustrated by the results obtained for the case of the Vector-Fermion Dark Matter (VFDM) model \cite{Ahmed:2017dbb}, which extends the SM gauge group by an~additional $U(1)_x$ with vector $X$ serving as its gauge boson. In order to provide mass to $X$, the Higgs mechanism with a~complex singlet $S$ is employed. Scalar~$S$ mixes with the SM Higgs doublet $H$ providing a~Higgs portal that communicates SM to the dark sector. Diagonalizing the mass-squared matrix of the real-part fluctuations of $S$ and $H$ one obtains two mass eigenstates: $h_1$ identified with the known Higgs particle of mass $125\gev$, and $h_2$ whose mass can have, in principle, any value. The model also introduces two Majorana fermions denoted as $\psi_+$ and $\psi_-$, coupled to $h_{1,2}$ via Yukawa interaction with the coupling constant $y_x\equiv g_x(m_{\psi_+}-m_{\psi_-})/(2m_X)$, where $g_x$ is the coupling constant of $U(1)_x$.

Particles $X$, $\psi_+$ and $\psi_-$, neutral under the action of the SM gauge group and charged under $U(1)_x$, can serve as dark matter candidates, interacting with the SM through a~$h_{1,2}$-mediated Higgs portal. Due to the presence of the $X\psi_+\psi_-$ interaction vertex, either two or three of those particles are stable, depending on whether the values of the masses allow one of them to decay into the two others.

The model parameters are masses of the potentially dark particles: $m_X$, $m_{\psi_+}$, $m_{\psi_-}$ (by definition, $m_{\psi_-}$ is always smaller than $m_{\psi_+}$), mass of the second Higgs state $m_{h_2}$, the~$U(1)_x$ interaction constant $g_x$, and the sine of the scalar-sector mixing angle $\sin\alpha$.

Values of $X_0$ calculated within the VFDM model are presented in the following \cref{tab:X0X,tab:X0p,tab:X0m}. A~given loop contributes to the self-energy if the first of loop particles is allowed to decay into the second one and the mediator. For details of calculation of the results shown in the tables, see \cref{app:X0table-general,app:X0table-VFDM}.

\begin{table}[H]
\begin{center}\begin{tabular}{|ccc|c|}
\hline
\multicolumn{3}{|c|}{mediator: $X$}	&	singular process\\
\hline
	&&&\multirow{8}{*}{
		\includegraphics[width=2.1cm]{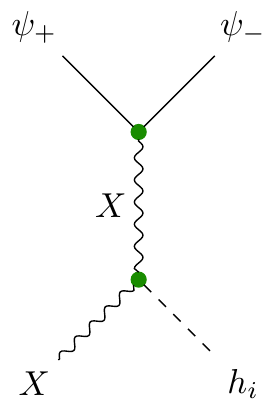}
	}\\
	loop states	&	$X_0$	& condition	&\\
	&&&\\
	$\psi_+\psi_-$	&	$\phantom{+}	\frac{1}{3}|\mathcal M|_{\psi_+\to X\psi_-}^2$
					&	$m_{\psi_+}>m_X+m_{\psi_-}$
					&	\\
	&&&\\&&&\\
	$h_iX\quad$		&	$-			\frac{1}{3}|\mathcal M|_{h_i\to XX}^2$
					&	$m_{h_i}>2m_X$
					&	\\
	&&&\\
\hline
\end{tabular}\end{center}
\caption{Values of the factor $X_0$ calculated within the VFDM model for $X$ being the mediator. Conditions for the loops to contribute to the effective width are also provided.}
\label{tab:X0X}
\end{table}

\begin{table}[H]
\begin{center}\begin{tabular}{|ccc|c|}
\hline
\multicolumn{3}{|c|}{mediator: $\psi_+$}		&	singular process\\
\hline
	&&&\multirow{8}{*}{
		\includegraphics[width=2.1cm]{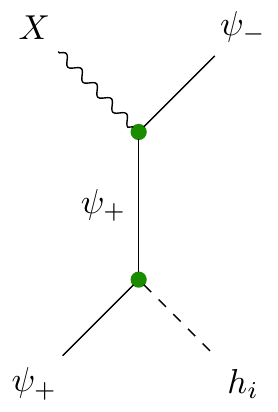}
	}\\
	loop states	&	$X_0$	&	condition	&\\
	&&&\\
	$X\psi_-$	&	$\phantom{+}	\frac{1}{2}|\mathcal M|_{X\to\psi_+\psi_-}^2$
				&	$m_X>m_{\psi_+}+m_{\psi_-}$
				&	\\
	&&&\\	&&&\\
	$h_i\psi_+$	&	$\phantom{+}	\frac{1}{2}|\mathcal M|_{h_i\to\psi_+\psi_+}^2$
				&	$m_{h_i}>2m_{\psi_+}$
				&	\\
	&&&\\
\hline
\end{tabular}\end{center}
\caption{Values of the factor $X_0$ calculated within the VFDM model for $\psi_+$ being the mediator. Conditions for the loops to contribute to the effective width are also provided.}
\label{tab:X0p}
\end{table}

\begin{table}[H]
\begin{center}\begin{tabular}{|ccc|c|}
\hline
\multicolumn{3}{|c|}{mediator: $\psi_-$}	&	singular process\\
\hline
	&&&	\multirow{9}{*}{
			\raisebox{-1.5cm}{			
			\raisebox{-0.5\height}{
				\includegraphics[width=2.1cm]{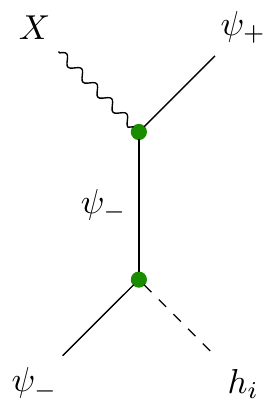}
			}
			or
			\raisebox{-0.5\height}{
				\includegraphics[width=2.1cm]{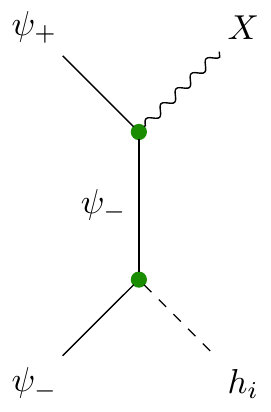}
			}
			}
		}\\
	loop states	&	$X_0$	&	condition	&\\
	&&&\\
	$X\psi_+$	&	$\phantom{+}	\frac{1}{2}|\mathcal M|_{X\to\psi_+\psi_-}^2$	
				&	$m_X>m_{\psi_+}+m_{\psi_-}$
				&	\\
	&&&\\
	$\psi_+X$	&	$-			\frac{1}{2}|\mathcal M|_{\psi_+\to X\psi_-}^2$
				&	$m_{\psi_+}>m_X+m_{\psi_-}$
				&	\\
	&&&\\
	$h_i\psi_-$	&	$\phantom{+}	\frac{1}{2}|\mathcal M|_{h_i\to\psi_-\psi_-}^2$
				&	$m_{h_i}>2m_{\psi_-}$
				&	\\
	&&&\\
\hline
\end{tabular}\end{center}
\caption{Values of the factor $X_0$ calculated within the VFDM model for $\psi_-$ being the mediator. Conditions for the loops to contribute to the effective width are also provided.}
\label{tab:X0m}
\end{table}

\begin{figure}[H]\begin{center}
	\includegraphics[width=0.49\textwidth]{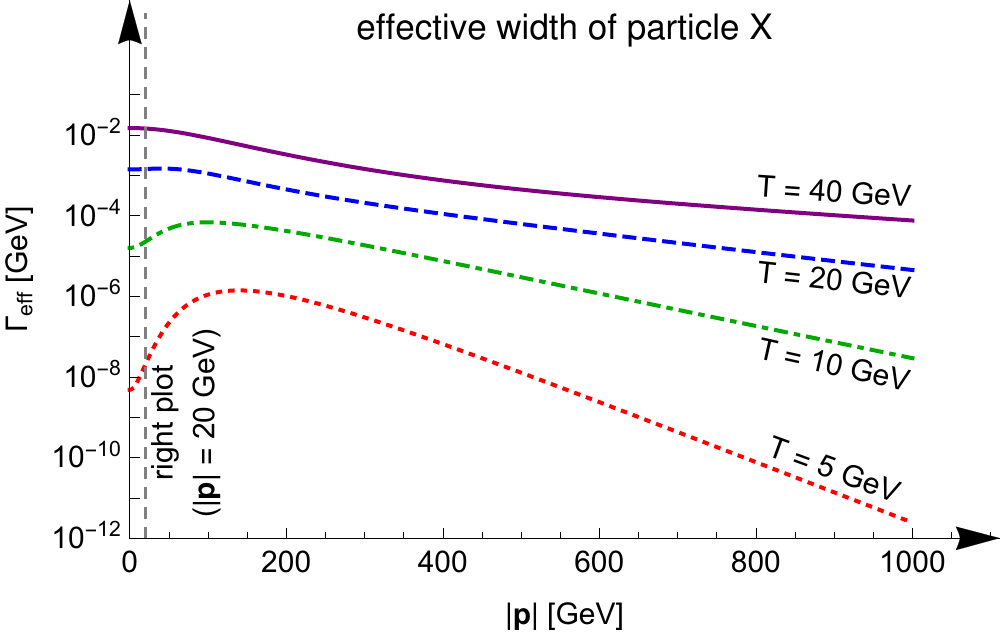}
	\includegraphics[width=0.49\textwidth]{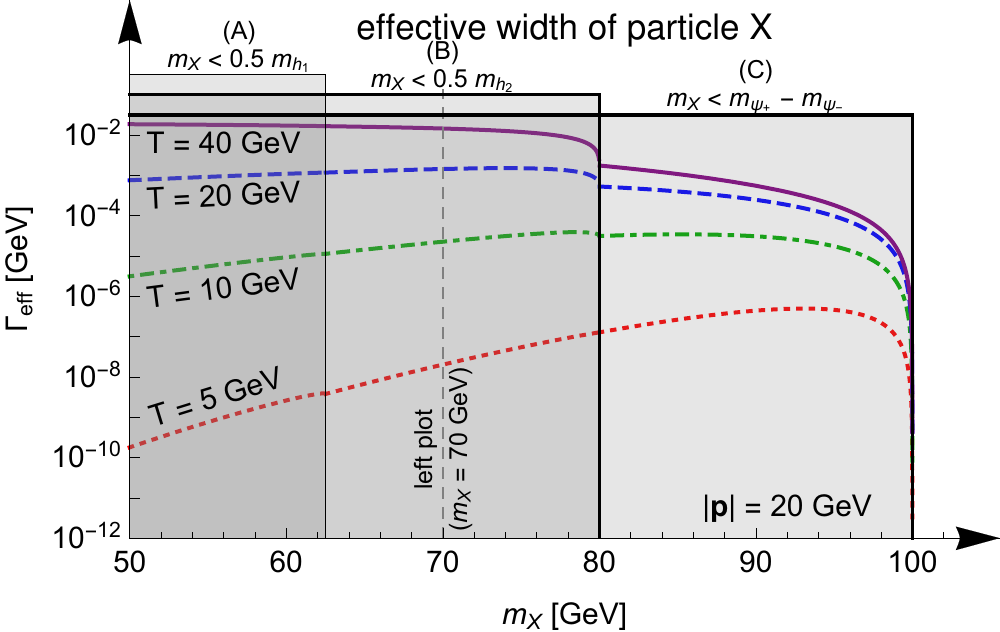}\\[5pt]
	\scalebox{0.6}{\begin{minipage}{0.45\textwidth}
		\quad\large\underline{values of the model parameters}
		\beq\notag
		m_X &= 70\gev\quad\mathrlap{\text{(left plot)}}\\
		m_{\psi_+} &= 130\gev	&	m_{h_1} &= 125\gev\\
		m_{\psi_-} &= 30\gev		&	m_{h_2} &= 160\gev\\
		g_x &= 0.1				&	\sin\alpha &= 0.1\\[14pt]
		\eeq
	\end{minipage}}%
	\begin{minipage}{0.05\textwidth}
	~
	\end{minipage}%
	\scalebox{0.6}{\begin{minipage}{\textwidth}
	\large\underline{contributing diagrams}\\[5pt]
	\includegraphics[width=0.3\linewidth]{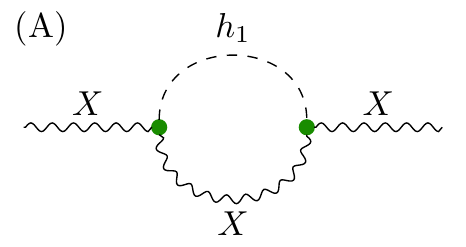}\hspace{0.5cm}
	\includegraphics[width=0.3\linewidth]{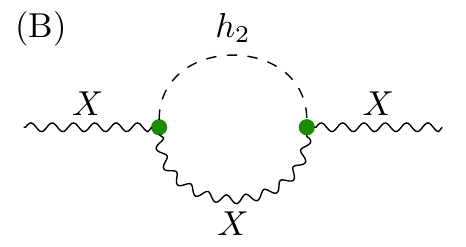}\hspace{0.5cm}
	\includegraphics[width=0.3\linewidth]{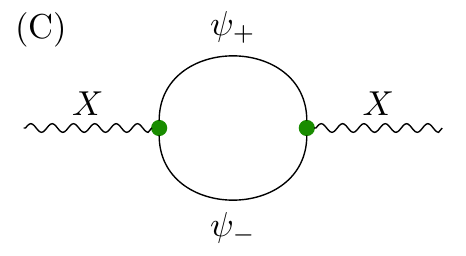}\hfill~
	\end{minipage}}
	\caption{
		Effective width of particle $X$, acquired as a~result of interactions with a~thermal medium
		of temperature $T=5\gev$ (dotted red), $T=10\gev$ (dot-dashed green), $T=20\gev$ (dashed blue), and $T=40\gev$ (solid purple),
		plotted as a~function of particle's momentum~$\n p$ ({\bf left}) and particle's mass~$m_X$ ({\bf right}).
		In each plot, the dashed vertical line indicates the value of the parameter corresponding to the other plot.
		In the {\bf right} plot, each gray rectangle (A), (B), (C) corresponds to the range of $m_x$ for which a~given loop (shown in the diagrams below the plots) contributes to the effective width.
		Used values of the model parameters are provided below the plots.
	}
	\label{fig:gammaX}
	\end{center}\end{figure}

\begin{figure}[H]\begin{center}
	\includegraphics[width=0.49\textwidth]{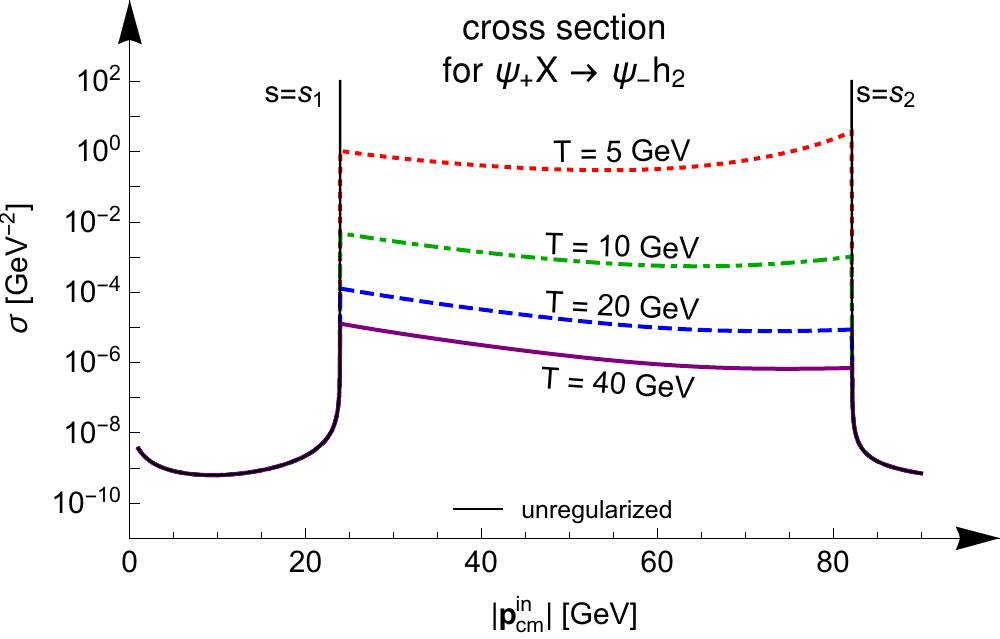}
		\begin{overpic}[width=0.49\textwidth]{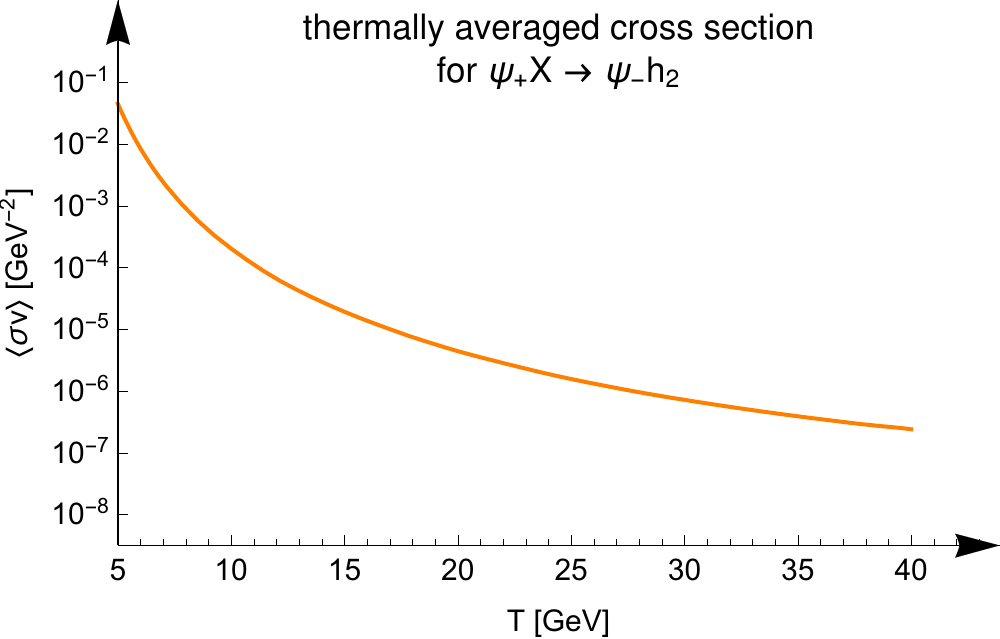}
		\put(74,23){\includegraphics[width=0.1\textwidth]{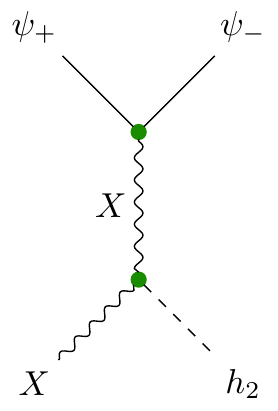}}
		\put(8,21){
			\scalebox{0.6}{\begin{minipage}{0.5\textwidth}
			\begin{center}
			\large\underline{values of}\kern150pt~\\[-2pt]
			\underline{model's parameters}\kern95pt~\\[-15pt]
			\beg\notag
			m_X = 70\gev			\quad	g_x = 0.1			\quad	\sin\alpha = 0.1	\\
			m_{\psi_+} = 130\gev	\quad	m_{h_1} = 125\gev							\\
			m_{\psi_-} = 30\gev	\quad	m_{h_2} = 160\gev
			\eeg
			\end{center}
			\end{minipage}}%
		}
		\end{overpic}
	\caption{
	{\bf Left:} regularized cross section for the~$\psi_+X\leftrightarrow\psi_-h_2$ process
	plotted as a~function of initial state CM momentum $|\vec p_\cm^\text{in}|$.
	The solid black line corresponds to the unregularized result (which is singular between $s=s_1$ and $s=s_2$),
	while the other lines show the regularized result for temperature of the medium equal to $T=5\gev$ (dotted red), $T=10\gev$ (dot-dashed green), $T=20\gev$ (dashed blue), and $T=40\gev$ (solid purple).
	Apparent non-smoothness of the coloured lines at $s=s_{1,2}$ is just an~effect of used scale.
	{\bf Right:}
Regularized thermally averaged cross section for the~$\psi_+X\leftrightarrow\psi_-h_2$ process as a~function of temperature of the medium $T$.
	Used values of model parameters (the same as in \cref{fig:gammaX}) and a~diagram of the considered process are provided.
	}
	\label{fig:sigmaPpX-PmH2}
	\end{center}\end{figure}

For an~illustration, the plots provided in \cref{fig:gammaX} show the values of $\Gamma_\eff$ calculated for particle $X$ being the mediator. \Cref{fig:sigmaPpX-PmH2} shows the cross section and the thermally averaged cross section calculated for the singular process ${\psi_+X\to\psi_-h_2}$ mediated by a~$t$-channel $X$, regularized by the effective width shown in \cref{fig:gammaX}.


\begin{figure}[H]\begin{center}
	\includegraphics[width=0.49\textwidth]{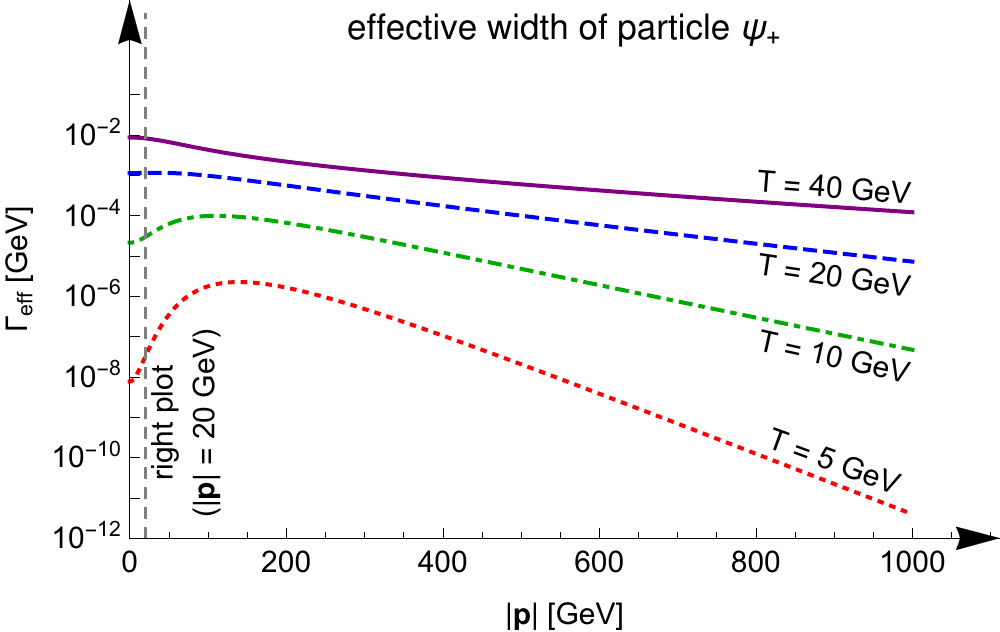}
	\includegraphics[width=0.49\textwidth]{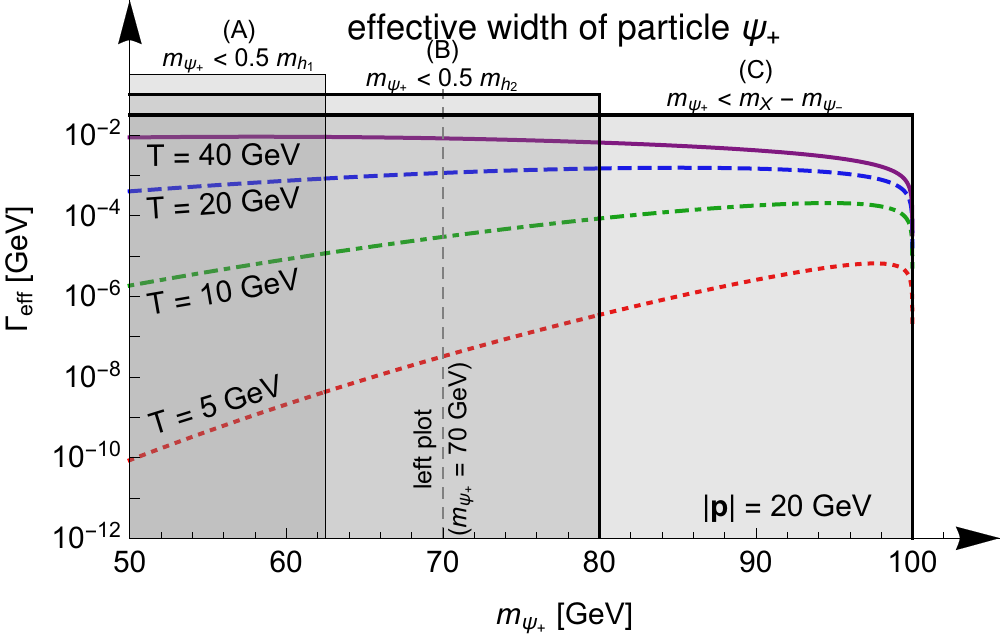}\\[5pt]
	\scalebox{0.6}{\begin{minipage}{0.5\textwidth}
		\quad\large\underline{values of model parameters}
		\beq\notag
		m_{\psi_+} &= 70\gev\quad\mathrlap{\text{(left plot)}}\\
		m_X &= 130\gev			&	m_{h_1} &= 125\gev\\
		m_{\psi_-} &= 30\gev		&	m_{h_2} &= 160\gev\\
		g_x &= 0.1				&	\sin\alpha &= 0.1\\[14pt]
		\eeq
	\end{minipage}}%
	\begin{minipage}{0.05\textwidth}
	~
	\end{minipage}%
	\scalebox{0.6}{\begin{minipage}{\textwidth}
	\large\underline{contributing diagrams}\\[5pt]
	\includegraphics[width=0.3\linewidth]{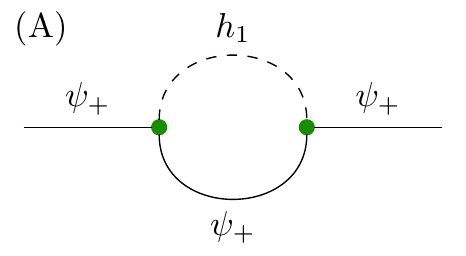}\hspace{0.5cm}
	\includegraphics[width=0.3\linewidth]{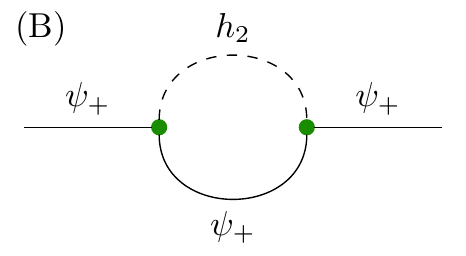}\hspace{0.5cm}
	\includegraphics[width=0.3\linewidth]{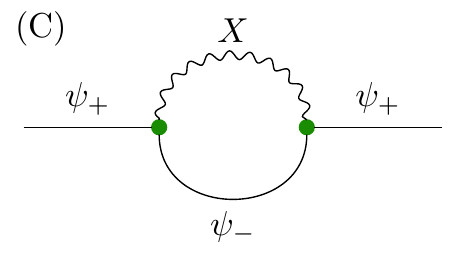}\hfill~
	\end{minipage}}
	\caption{
		Effective width of particle $\psi_+$, acquired as a~result of interactions with a~thermal medium
		of temperature $T=5\gev$ (dotted red), $T=10\gev$ (dot-dashed green), $T=20\gev$ (dashed blue), and $T=40\gev$ (solid purple),
		plotted as a~function of particle's momentum~$\n p$ ({\bf left}) and particle's mass~$m_{\psi_+}$ ({\bf right}).
		In each plot, the dashed vertical line indicate the value of the parameter corresponding to the other plot.
		In the {\bf right} plot, each gray rectangle (A), (B), (C) corresponds to the range of $m_{\psi_+}$ for which a~given loop (shown in the diagrams below the plots) contributes to the effective width.
		Used values of model parameters are provided below the plots.
	}
	\label{fig:gammaPp}
	\end{center}\end{figure}

\begin{figure}[H]\begin{center}
	\includegraphics[width=0.49\textwidth]{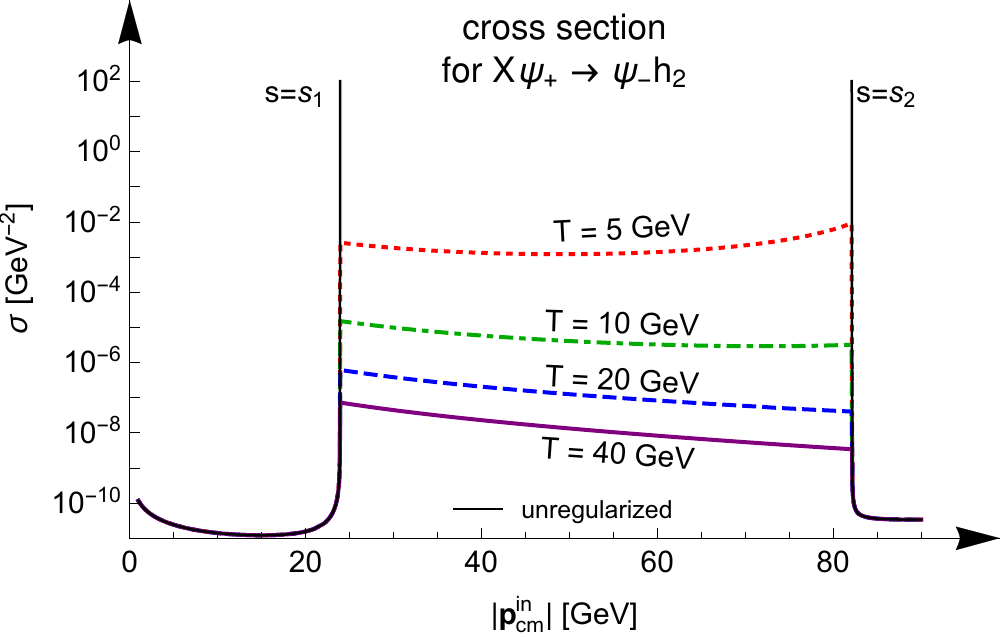}
		\begin{overpic}[width=0.49\textwidth]{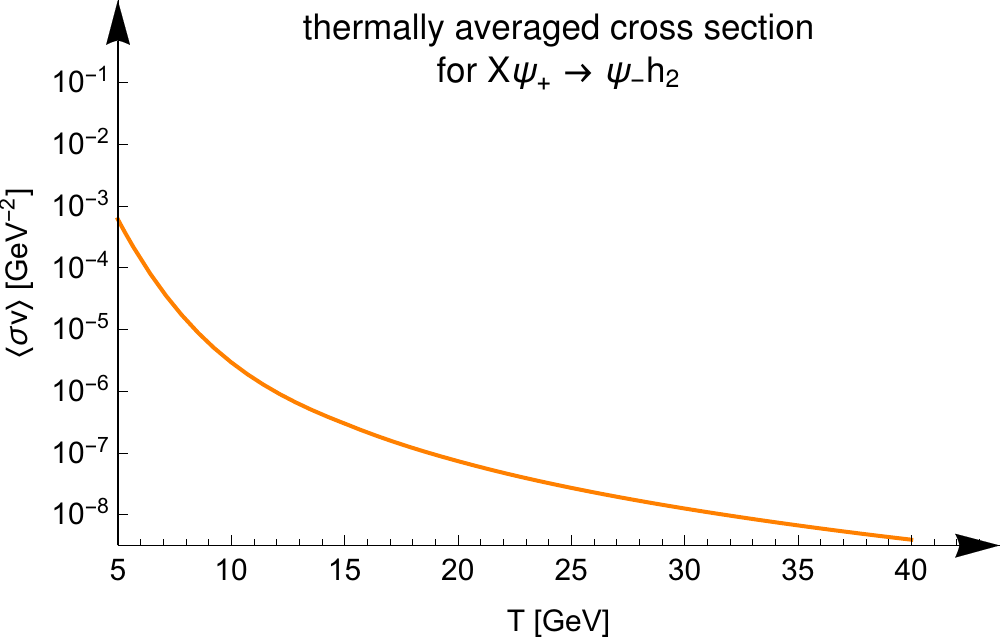}
		\put(74,18){\includegraphics[width=0.1\textwidth]{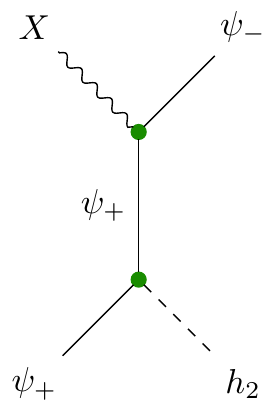}}
		\put(13,40){
			\scalebox{0.6}{\begin{minipage}{0.5\textwidth}
			\begin{center}
			\large\underline{values of model parameters}
			\beg\notag
			m_{\psi_+} = 70\gev	\quad	g_x = 0.1			\quad	\sin\alpha = 0.1	\\
			m_X = 130\gev		\quad	m_{h_1} = 125\gev							\\
			m_{\psi_-} = 30\gev	\quad	m_{h_2} = 160\gev
			\eeg
			\end{center}
			\end{minipage}}%
		}
		\end{overpic}
	\caption{
	{\bf Left:} regularized cross section for the~$X\psi_+\leftrightarrow\psi_-h_2$ process
	plotted as a~function of initial state CM momentum $|\vec p_\cm^\text{in}|$.
	The solid black line corresponds to the unregularized result (which is singular between $s=s_1$ and $s=s_2$),
	while the other lines show the regularized result for temperature of the medium equal to $T=5\gev$ (dotted red), $T=10\gev$ (dot-dashed green), $T=20\gev$ (dashed blue), and $T=40\gev$ (solid purple).
	Apparent non-smoothness of the coloured lines at $s=s_{1,2}$ is just an~effect of used scale.
	{\bf Right:}
Regularized thermally averaged cross section for the~$X\psi_+\leftrightarrow\psi_-h_2$ process as a~function of temperature of the medium $T$.
	Used values of model parameters (the same as in \cref{fig:gammaPp}) and a~diagram of the considered process are provided.
	}
	\label{fig:sigmaXPp-PmH2}
	\end{center}\end{figure}

Similarly, the plots provided in \cref{fig:gammaPp} show the values of $\Gamma_\eff$ calculated for particle $\psi_+$ being the mediator. \Cref{fig:sigmaXPp-PmH2} shows the cross section and the thermally averaged cross section calculated for the singular process ${X\psi_+\to\psi_-h_2}$ mediated by a~$t$-channel $\psi_+$, regularized by the effective width shown in \cref{fig:gammaX}.

\Cref{fig:gammaPp,fig:sigmaXPp-PmH2} are, in some sense, symmetric to \cref{fig:gammaX,fig:sigmaPpX-PmH2}. The difference is that the masses of particles $X$ and $\psi_+$ used to prepare the second set are swapped with respect to the first one, making it possible to compare processes of the same kinematics, but different spin of the mediating particle: $\nicefrac12$ for \cref{fig:gammaPp,fig:sigmaXPp-PmH2}, $1$ for \cref{fig:gammaX,fig:sigmaPpX-PmH2}.

Note that, in the notation assumed here, the processes denoted ${\psi_+X\to\psi_-h_2}$ and ${X\psi_+\to\psi_-h_2}$, although sharing the same initial and final states, are given by two different diagrams shown in \cref{fig:sigmaPpX-PmH2,fig:sigmaXPp-PmH2}, respectively. For the first of them, it is necessary for the singularity to occur that $m_{\psi_+} > m_X + m_{\psi_-}$, while for the second one, $m_X > m_{\psi_+} + m_{\psi_-}$ must hold.


\section{Summary}
\label{sec:summary}
In this paper, the method of regularization of the $t$-channel singularity for processes occurring in a~thermal medium, presented in \cite{Grzadkowski:2021kgi}, has been generalized to the case of particles of arbitrary spins (scalars, fermions and vectors). In \cref{sec:selfE} an~analytical result for each spin case is provided.

The singularity, which would lead to infinite cross section of the processes affected, is cured by introducing an~effective thermal width of the mediator. That width prevents the resummed propagator from being singular. Physically, the singularity, corresponding to would-be infinite lifetime (equivalently: free path) of the on-shell mediator, does not occur in a~thermal medium because interactions between the mediator and the surrounding particles provide a~channel for the mediator to vanish, making its lifetime (free path) finite.

The results presented in this paper appear to behave in a~natural, expected way: the effective thermal width vanishes for small temperatures (in the limit equivalent to lack of the medium) while in finite temperature it is non-trivial even for zero mediator's momentum, see \cref{sec:discussion}.

To illustrate the method in action, theoretical and numerical results calculated within the Vector-Fermion Dark Matter model \cite{Ahmed:2017dbb} have been presented in \cref{sec:X0VFDM}. The plots that are shown in that section follow the behaviour predicted in \cref{sec:discussion}.

It is worth to stress that this work shows the full derivation of the method and demonstrates an~application to an~actual, specific case. This makes the presented results easy to reproduce and adapt to reader's own research.

\acknowledgments
I am grateful to B.~Grzadkowski for encouraging me to write this paper and for discussions concerning it. I would also like to thank K{\'a}roly Seller for a~fruitful discussion and bringing my attention to the papers \cite{Dicus:1982bz,Weldon:1983jn,Giudice:2003jh}. I thank St.~Mr{\'o}wczy{\'n}ski for enabling me to familiarize myself with \cite{Bellac:2011kqa}.

This work has been partially supported by the National Science Centre (Poland) under grants 2017/25/B/ST2/00191 and 2020/37/B/ST2/02746. 

\appendix

\section{Conventions}
This paper assumes the ,,mostly plus'' metric tensor:
	\beq
	g_{\mu\nu}=\text{diag}(+,-,-,-)\;.
	\eeq
Feynman diagrams are calculated in the unitary gauge. The Feynman propagator of a~ scalar, a~fermion and a~vector particle, respectively, is given by
	\beq
	&\includegraphics[scale=1]{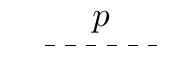}
		&&\to & i\,\Delta(p)&\equiv\frac{i}{p^2-m^2}\;,\\
	&\includegraphics[scale=1]{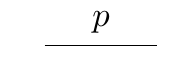}
		&&\to & i\,G(p)&\equiv\frac{i}{p^2-m^2}\,(\sl p+m)\;,\\
	&\includegraphics[scale=1]{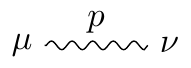}
		&&\to & i\,D_{\mu\nu}(p)&\equiv\frac{i}{p^2-m^2}
			\left(-g_{\mu\nu}+\frac{p_\mu p_\nu}{m^2}\right)\;.
	\eeq

\section{Green's functions}
\label{app:green}
Here, following \cite{Czajka:2010zh}, the retarded (,,$+$''), advanced (,,$-$''), and symmetric (,,$\sym$'') propagators of a~scalar, fermion, and vector particle are provided. For a~scalar particle, in the position space, those propagators are related to the real-time Green functions $\Delta^>$, $\Delta^<$ in the following way:
	\beq
	\Delta^\sym(x,y)	&= \Delta^>(x,y) + \Delta^<(x,y)\\
	\Delta^+(x,y)	&= \Theta(x_0-y_0)\left(\Delta^>(x,y) - \Delta^<(x,y)\right)\\
	\Delta^-(x,y)	&= \Theta(y_0-x_0)\left(\Delta^<(x,y) - \Delta^>(x,y)\right)\;,
	\eeq
where $\Theta(t-t')$ is equal to 1 if $t$ succeeds $t'$ along the integration contour of the Keldysh-Schwinger formalism, $-1$ otherwise. Analogous relations hold for fermion and vector propagators. Details can be found in, e.g., section 4.1 of \cite{Czajka:2015mvx}.
 
The propagators will be expressed in terms of the auxiliary functions $\Delta^{\pm}_\text{aux}$, $\Delta_\text{F,B}^\sym(p,T)$:
	\beq
	\Delta^{\pm}_\text{aux}(p)
	&\equiv
		\frac{1}{p^2 - m^2 \pm i\, \sgn p_0\,0^+}\,,\\
	\Delta_\text{F,B}^\sym(p,T)
	&\equiv
		\pm\frac{i\pi}{E_p}
		\Big( \delta (E_p - p_0) \big[2 n_\text{F,B}(\vec p,T) \mp 1\big]
	+	\delta (E_p + p_0) \big[2 n_\text{F,B}(-\vec p,T) \mp 1\big] \Big)\;,
	\eeq
where $n_\text{F,B}$ denotes the momentum-space distribution function of the considered kind of particles, with index F standing for fermions and B for bosons. Since in equilibrium\footnote{The assumption of equilibrium is convenient but not necessary, see \cref{ft:T}.}
	\beq
	n_\text{F,B}(-\vec p,T)=n_\text{F,B}(\vec p,T)=\frac{1}{e^{\beta E_p}\pm1}\;,
	\eeq
the $\Delta_\text{F,B}^\sym$ propagator is given by
	\beq
	\Delta_{\text{F},\text{B}}^\sym(p,T)
	&\equiv
	\pm	\frac{i\pi}{E_p}
		\Big( \delta (E_p - p_0)	+  \delta (E_p + p_0)\Big)
		\big[2 n_{\text{F},\text{B}}(\vec p,T) \mp 1 \big]\\
	&=
	-	\frac{i\pi}{E_p}
		\Big( \delta (E_p - p_0)	+  \delta (E_p + p_0)\Big)
		\cdot f_{\text{F,B}}(\beta E_p)\;,
	\eeq
where
	\beq
	f_{\text{F,B}}(x)
	&\equiv
		\begin{cases}
			\frac{e^x-1}{e^x+1}	&\text{ for fermions}\\
			\frac{e^x+1}{e^x-1}	&\text{ for bosons}
		\end{cases}\;.
	\eeq
For a~scalar field, the retarded, advanced, and symmetric propagators are given by
	\beq\label{eq:propSca}
	\Delta^{\pm}(p) &= \Delta^{\pm}_\text{aux}(p),\\
	\Delta^\sym(p,T) &= \Delta_B^\sym(p,T)\;,
	\eeq
while for a fermion
	\beq\label{eq:propFer}
	G^\pm(p) &=(\sl p+m)\,\Delta^\pm_\text{aux}(p)\;,\\
	G^\sym(p,T) &=(\sl p+m)\,\Delta_F^\sym(p,T)\;,
	\eeq
and for a vector
	\beq\label{eq:propVec}
	D_{\mu\nu}^\pm(p) &=\left[-g_{\mu\nu}+\frac{p_\mu p_\nu}{m^2}\right]\,\Delta^\pm_\text{aux}(p)\;,\\
	D_{\mu\nu}^\sym(p,T) &=\left[-g_{\mu\nu}+\frac{p_\mu p_\nu}{m^2}\right]\,\Delta^\sym_B(p,T)\;.
	\eeq

\section{Allowed range of \topstr{$E_{1,2}$}{E\_1,2}}
\label{app:krange}
In this appendix, a~proof of correctness of the transition between \cref{eq:selfE1} and \cref{eq:deltacos} is presented. Let us recall the definitions of $\cos\alpha_1$, $\cos\beta_1$, $\cos\alpha_2$, and $\cos\beta_2$ given by \cref{eq:cos}:

	\beq
	\cos\alpha_1
		&\equiv \frac{-(m_1^2 - m_2^2 + M^2) + 2E_1E_p}{2\n k\n p}\;,&
	\cos\beta_1
		&\equiv \frac{-(m_1^2 - m_2^2 + M^2) - 2E_1E_p}{2\n k\n p}\;,\\
	\cos\alpha_2
		&\equiv \frac{-(m_1^2 - m_2^2 - M^2) + 2E_2E_p}{2\n k\n p}\;,&
	\cos\beta_2
		&\equiv \frac{-(m_1^2 - m_2^2 - M^2) - 2E_2E_p}{2\n k\n p}\;.
	\eeq
If $m_2 > |m_1 - M|$, then $\cos\alpha_1 > 1$. Indeed,
	\beq
	\cos\alpha_1
	= \frac{-m_1^2 + m_2^2 - M^2 + 2E_1E_p}{2\,\n{k}\n{p}}
	> \frac{-2m_1M + 2E_1E_p}{2\,\n{k}\n{p}}
	> 1\;,
	\eeq
Analogously, if $m_1 > |m_2 - M|$, then $\cos\beta_2 < -1$, since
	\beq
	\cos\beta_2
	= -\frac{m_1^2 - m_2^2 - M^2 + 2E_2E_p}{2\,\n{k}\n{p}}
	< -\frac{-2m_2M + 2E_2E_p}{2\,\n{k}\n{p}}
	< -1\;.
	\eeq
Moreover, if $m_2 < m_1 + M$, then $\cos\beta_1 < -1$:
	\beq
	\cos\beta_1
	= -\frac{m_1^2 - m_2^2 + M^2 + 2E_1E_p}{2\,\n{k}\n{p}}
	< -\frac{-2m_1M + 2E_1E_p}{2\,\n{k}\n{p}}
	< -1\;.
	\eeq
Finally, if $m_1 < m_2 + M$, then $\cos\alpha_2 > 1$:
	\beq
	\cos\alpha_2
	= \frac{-m_1^2 + m_2^2 + M^2 + 2E_2E_p}{2\,\n{k}\n{p}}
	> \frac{-2m_2M + 2E_2E_p}{2\,\n{k}\n{p}}
	> 1\;,
	\eeq
Together, these four statements mean that if $m_1$, $m_2$ and $M$ satisfy the triangle inequality, all four trigonometric functions are outside the $[-1,1]$ interval. Therefore, for the result of \cref{eq:selfE1} to be non-zero, one of the masses must be greater than the two others. Due to the assumption of stability of the mediator, $M$ cannot be greater than the sum of $m_1$ and $m_2$ (see \cref{ineq:stable}). This means that only the loops with one internal state allowed to decay into the other one and the mediator provide a non-zero contribution to the regulator.%
\footnote{
	Such a loop always exists because for the considered $t$-channel process to be singular it is necessary that the mediator can be produced in a decay process, see \cref{cond:sing}. Consequently, any $t$-channel singular $2\to 2$ process occurring in a medium can be regularized using the method described in this paper.
}
Without loss of generality it can be, therefore, assumed that
	\beq	\label{eq:tchannel:massCond}
	m_1 > m_2 + M\;,
	\eeq
so that $\cos\beta_1$ and $\cos\beta_2$ are outside the $[-1,1]$ interval and the delta functions containing them vanish.

Concerning the delta function containing $\cos\alpha_1$, it limits possible values of $E_1$ to those satisfying $|\cos\alpha_1|<1$. It appears that
	\beq\label{eq:parabola}
	k^2p^2(\cos^2\!\alpha_1 - 1) =
	M^2\,E_1^2 - (m_1^2 - m_2^2 + M^2)\,E_p\,E_1 + m_1^2\,E_p^2 + \frac{\lambda(m_1^2,m_2^2,M^2)}{4}\;,
	\eeq
where $\lambda$ denotes the so-called K\"all\'en function, defined as
	\beq\label{eq:lambda}
	\lambda(m_1^2,m_2^2,M^2)&\equiv m_1^4+m_2^4+M^4-2\,m_1^2m_2^2-2\,m_1^2M^2-2\,m_2^2M^2\\
	&= \left[m_1^2-(m_2-M)^2\right]\left[m_1^2-(m_2+M)^2\right]\;.
	\eeq
$E_1$ is in the allowed range if the quantity \labelcref{eq:parabola} is negative, which can be reduced to the following condition:
	\beq\label[ineq]{eq:e1}
	b-a < E_1 < b+a\;,
	\eeq
where
	\beg
	\cb\equiv\frac{m_1^2-m_2^2+M^2}{2M^2}\,E_p\;,\qquad
	\ca\equiv\frac{\lambda(m_1^2,m_2^2,M^2)^{1/2}}{2M^2}\,\n p\;.
	\eeg
Analogously, it can be shown that the delta function containing $\cos\alpha_2$ limits the range of $E_2$ to
	\beq\label[ineq]{eq:e2}
	b-a-E_p < E_2 < b+a-E_p\;.
	\eeq
Note that
	\beq\label{eq:positive}
	\cb -\ca -E_p
	=\frac{m_1^2-m_2^2-M^2}{2M^2}\,E_p - \frac{\lambda(m_1^2,m_2^2,M^2)^{1/2}}{2M^2}\,\n p
	\eeq
is positive since
	\beg
	m_1^2-m_2^2-M^2>\sqrt{(m_1^2-m_2^2-M^2)^2-4\,m_2^2\,M^2}=\lambda(m_1^2,m_2^2,M^2)^{1/2}
	\eeg
and
	\beg
	E_p>\n p\;.
	\eeg

\section{Calculation of \topstr{$X_0$}{X0} in the general case}
\label{app:X0table-general}
Here, the factor $X_0$ is calculated for all loops that can be generated using particles of spin $0$, $\nicefrac12$ and $1$. \Cref{tab:X0-general-sca,tab:X0-general-fer,tab:X0-general-vec} show the results for the case of a scalar, a fermion and a vector mediator, respectively. 

	\begin{table}[H]
	\includegraphics[width=\textwidth]{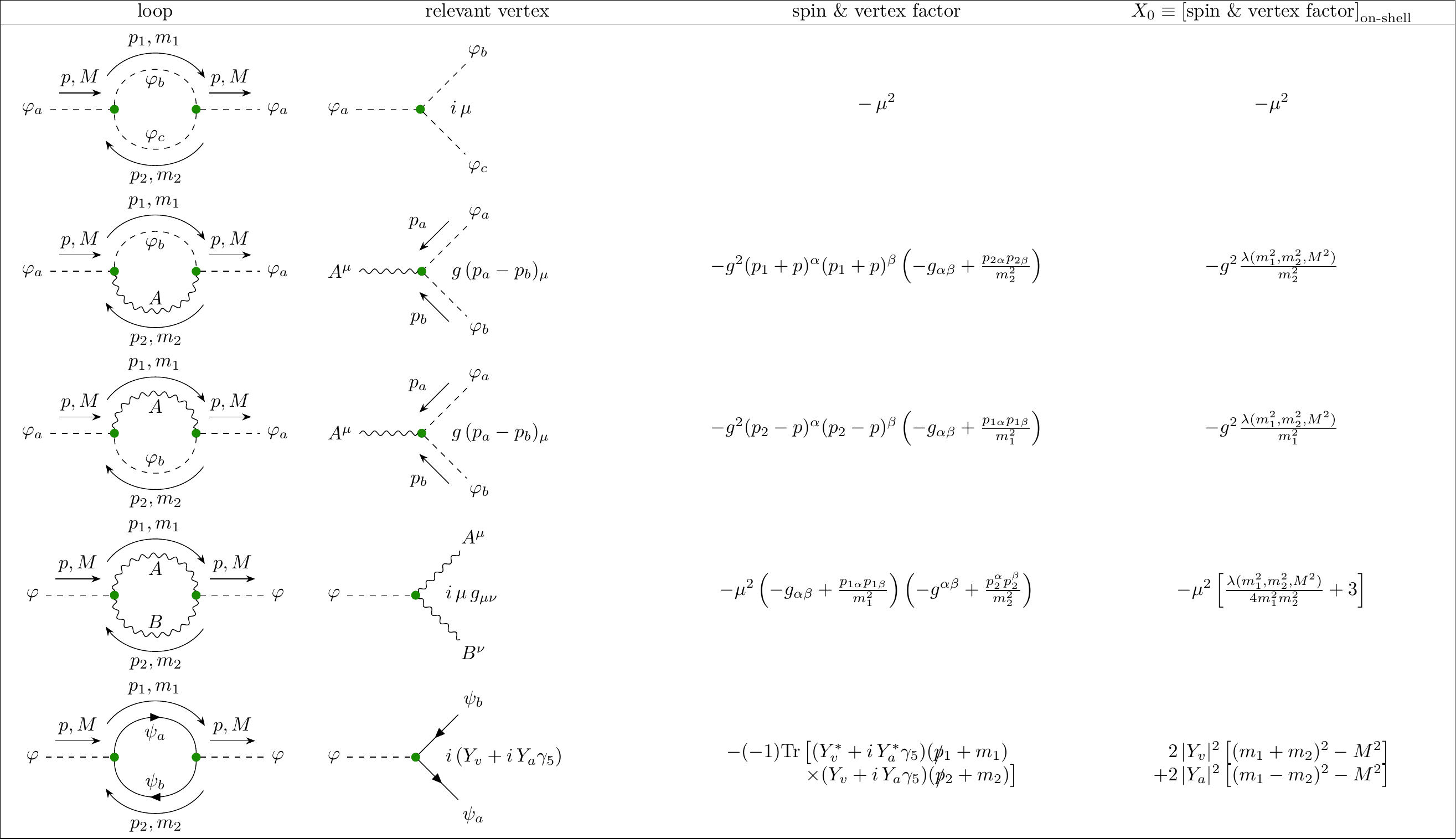}
	\caption{Factor $X_0$ calculated for all loops relevant in the case of a~scalar mediator. In the third column, ,,spin \& vertex factor'' denotes the matrix element corresponding to the loop, multiplied by $(-i)^2\left[p_1^2-m_1^2\right]\left[p_2^2-m_2^2\right]$, before integration over loop momenta. In the fourth column, the factor $X_0$ is calculated assuming that all particles are on-mass-shell. Note that $X_0$ is always positive (negative) for particle ,,2'' being a~fermion (boson) given that $m_1>m_2+M$. Scalar and vector states are assumed to be real (not complex). Here, $g$ is real and dimensionless, $\mu$ is real and has dimension of mass, while $Y_v$ and $Y_a$ are complex and dimensionless.}
	\label{tab:X0-general-sca}
	\end{table}
	
	\begin{table}[H]
	\includegraphics[width=\textwidth]{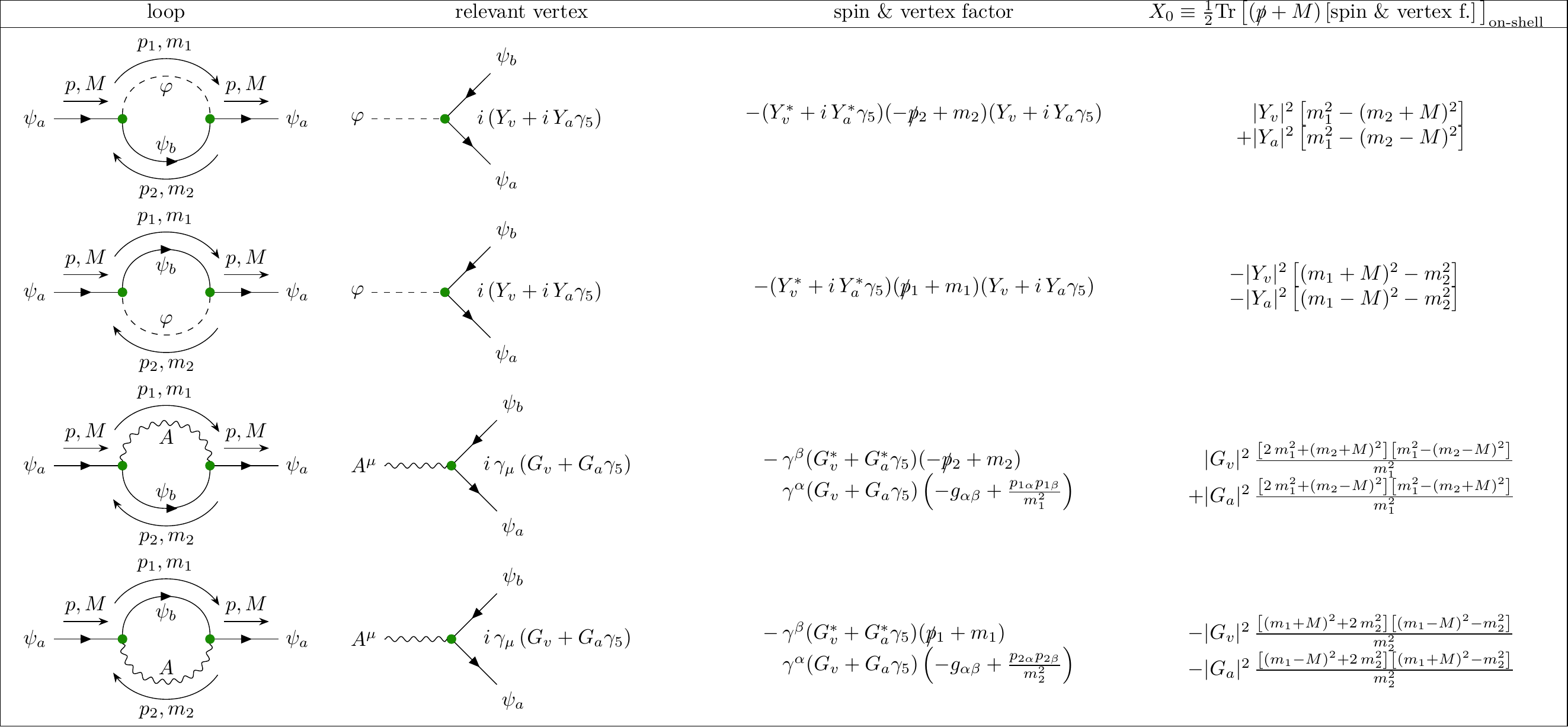}
	\caption{Like \cref{tab:X0-general-sca}, but for a~fermion mediator. Here, $Y_v$, $Y_a$, $G_v$, and $G_a$ are complex and dimensionless.}
	\label{tab:X0-general-fer}
	\end{table}
	
	\begin{table}[H]
	\includegraphics[width=\textwidth]{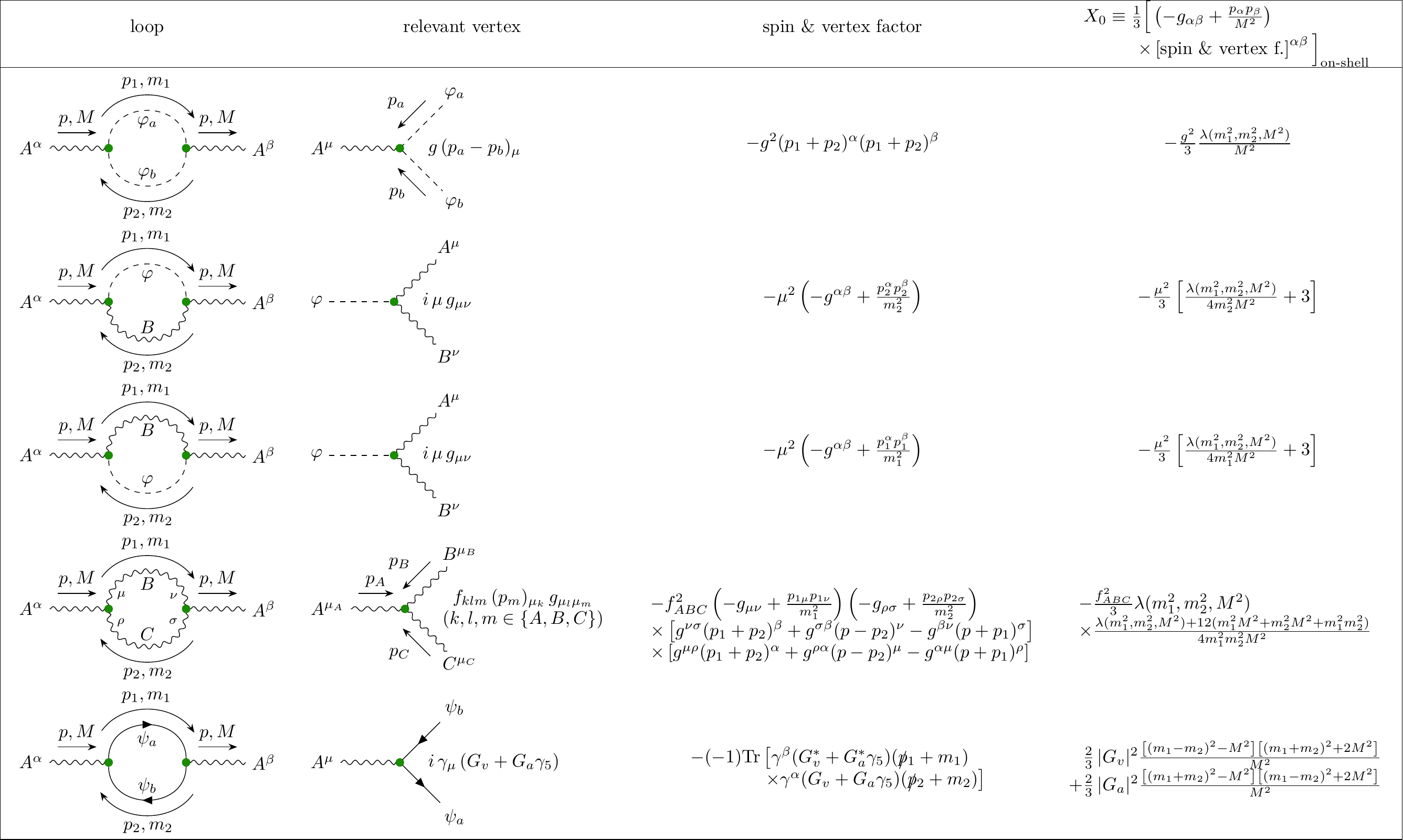}
	\caption{Like \cref{tab:X0-general-sca}, but for a~vector mediator. Here, $g$ and $f_{klm}$ are real and dimensionless, $\mu$ is real and has dimension of mass, while $G_v$ and $G_a$ are complex and dimensionless. It is assumed that $f_{klm}$ is totally antisymmetric in its indices.}
	\label{tab:X0-general-vec}
	\end{table}

\section{Calculation of \topstr{$X_0$}{X0} within the VFDM model}
\label{app:X0table-VFDM}
\begin{figure}[H]\begin{center}
	\includegraphics[scale=0.75]{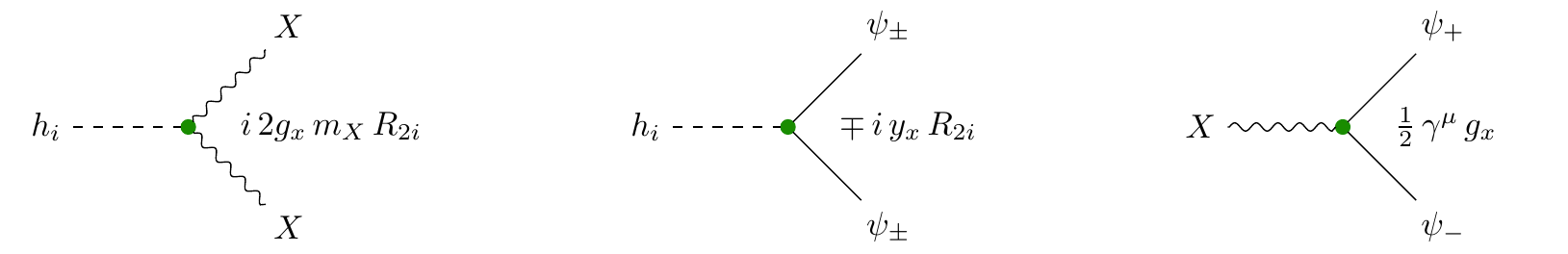}
	\caption{Relevant vertices of the VFDM model and their values.}
	\label{fig:vertices}
	\end{center}\end{figure}

\begin{figure}[H]\begin{center}
	\includegraphics[scale=1]{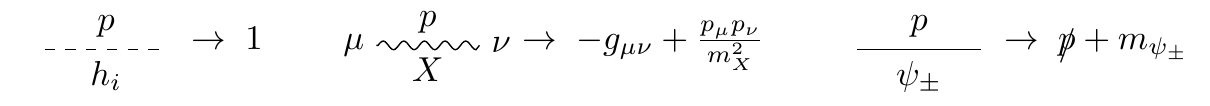}
	\caption{Spin factors (i.e., propagators with the factor $i\,[p^2-m^2\pm i\,0^+]^{-1}$ omitted) corresponding to the relevant particles of the model.}
	\label{fig:spinFactors}
	\end{center}\end{figure}

In this appendix, the values of the factor $X_0$ calculated for dark mediators present in the VFDM model (see \cref{sec:X0VFDM}) are presented. \Cref{fig:vertices} shows the relevant vertices while \cref{fig:spinFactors} provides the spin-dependent factors needed to calculate $X_0$. \Cref{tab:X0X-loops,tab:X0p-loops,tab:X0m-loops} present the values of $X_0$ for all contributing loops, basing on the results of \cref{app:X0table-general}.

	\begin{table}[H]\begin{center}
	\includegraphics[width=\textwidth]{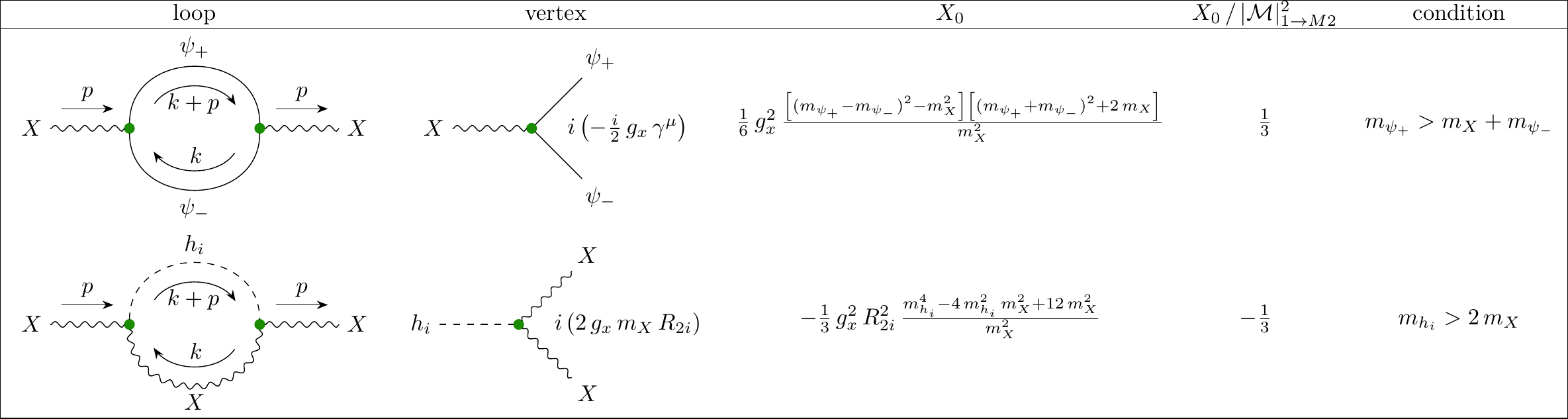}
	\caption{Factor $X_0$ calculated within the VFDM model for particle $X$ being the mediator. The last but one column contains the ratio between $X_0$ and the matrix-element-squared corresponding to the decay of the upper loop state into the lower loop state and the mediator (symbols ,,1'', ,,2'' and ,,$M$'' correspond to those in \cref{fig:loop}). The last column presents the condition for the given loop to contribute to the mediator's thermal width.}
	\label{tab:X0X-loops}
	\end{center}\end{table}
	\begin{table}[H]\begin{center}
	\includegraphics[width=\textwidth]{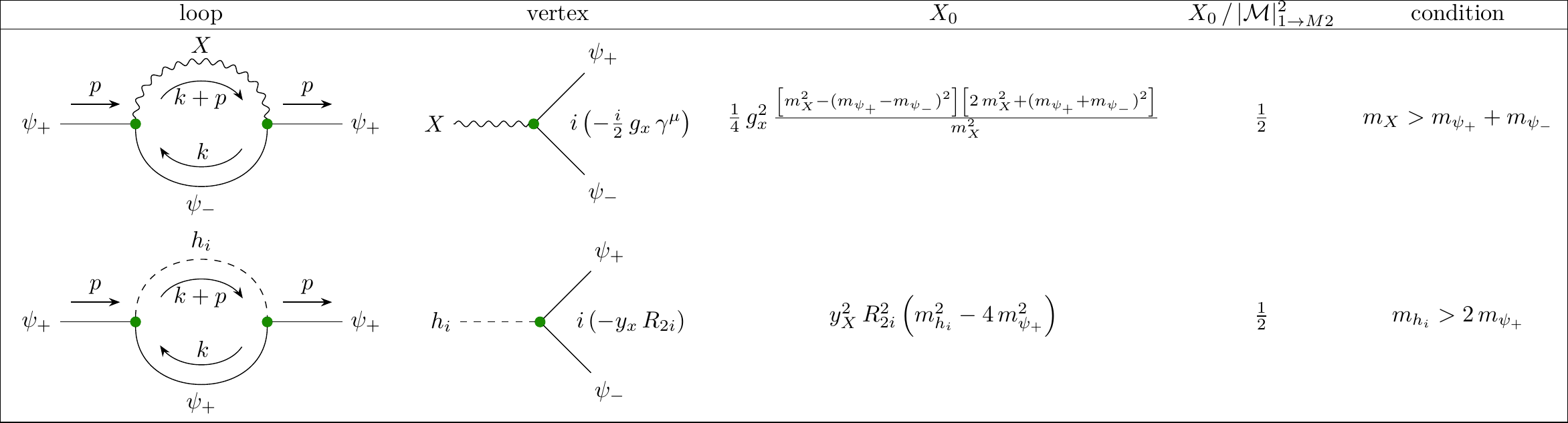}
	\caption{Like \cref{tab:X0X}, but for particle $\psi_+$ being the mediator.}
	\label{tab:X0p-loops}
	\end{center}\end{table}
	\begin{table}[H]\begin{center}
	\includegraphics[width=\textwidth]{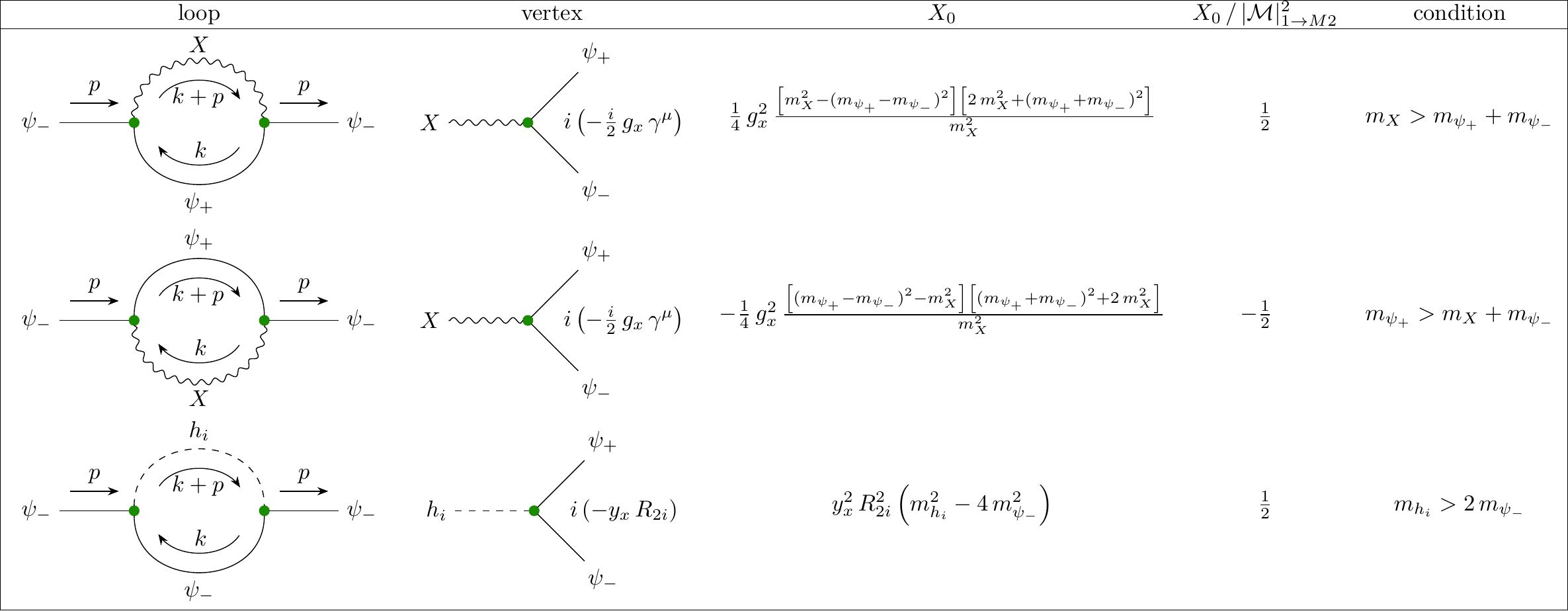}
	\caption{Like \cref{tab:X0X}, but for particle $\psi_-$ being the mediator.}
	\label{tab:X0m-loops}
	\end{center}\end{table}

\section{Results of Weldon's paper}
\label{app:weldon}
According to Weldon's paper \cite{Weldon:1983jn}, the imaginary part of thermal self-energy corresponding to a loop consisting of particles ,,$1$'' and ,,$2$'' (see \cref{fig:loop}) can be expressed as:
	\beq
	\Sigma
	&=\frac{1}{2}\int
		\frac{d^3p_1}{(2\pi)^32E_1}\frac{d^3p_2}{(2\pi)^32E_2}\,
		(2\pi)^4\delta^{(4)}(p_1-p_2-p)\,
		|\mathcal{M}_{1\to 2,M}|^2\,
		\left[\eta\, n_1(\beta E_1) + n_2(\beta E_2)\right]\;,
	\eeq
where
	\beg
	\eta\equiv
	\begin{cases}
		-1	&\text{if particles $1$ and $2$ are both bosons (both fermions)}\\
		+1	&\text{otherwise}
	\end{cases}\;,
	\eeg
while $n_i(x)\equiv\left(e^x\pm 1\right)^{-1}$, $i=1,2$ is the thermal distribution function, with the $\pm$ sign being a plus if particle $i$ is a fermion and a minus for a boson. After integration over $d^3p_2$ and a change of variables, the integral reads
	\beq
	\Sigma
	&=\frac{1}{16\pi\n p}
		\int_{m_1}^\infty dE_1
		\int_{-1}^1 d\cos\theta\,
		\delta(\cos\theta-\cos\theta_0)\,
		|\mathcal{M}_{1\to 2,M}|^2\,
		\left[\eta\, n_1(\beta E_1) + n_2(\beta E_2)\right]\;,
	\eeq
where
	\beq
	\cos\theta_0(E_1)\equiv\frac{2\,E_pE_1-(m_1^2-m_2^2+M^2)}{2\,\n p\n{p_1}}
	\eeq
is the value of $\cos\theta$ ($\theta$ denotes the angle between $\vec p$ and $\vec p_1$) corresponding to $E_1=E_2+E_p$. The delta function limits the range of integration over $E_1$ to the values giving $|\cos\theta_0|<1$:
	\beq
	\Sigma
	&=\frac{1}{16\pi\n p}\int_{E_-}^{E_+}
		dE_1\,
		|\mathcal{M}_{1\to 2,M}|^2\,
		\left[\eta\, n_1(\beta E_1) + n_2(\beta E_1-\beta E_p)\right]\;,
	\eeq
where
	\beg
	E_{\pm}\equiv\cb \pm\ca\;,\qquad
	\ca\equiv\frac{\n p\,\sqrt{\lambda(m_1^2,m_2^2,M^2)}}{2M^2}\;,\qquad
	\cb\equiv\frac{m_1^2-m_2^2+M^2}{2M^2}\,E_p\;,\\
	\lambda(m_1^2,m_2^2,M^2)\equiv\left[m_1^2-(m_2-M)^2\right]\left[m_1^2-(m_2+M)^2\right]\;.
	\eeg
Then, assuming Lorentz invariance of $|\mathcal M_{1\to 2,M}|^2$, the result can be transformed to
	\beq
	\Sigma=\Sigma(E_p,T)
	&=\frac{|\mathcal{M}_{1\to 2,M}|^2}{16\pi\n p}
		\int_{E_-}^{E_+}
		dE_1\,
		\left[\eta\, n_1(\beta E_1) + n_2(\beta E_1-\beta E_p)\right]\\
	&=\frac{|\mathcal{M}_{1\to 2,M}|^2}{16\pi\beta\n p}
		\left[
		\eta
		\int_{\beta E_-}^{\beta E_+}
		n_1(x)\,dx
		+
		\int_{\beta( E_--E_p)}^{\beta(E_+-E_p)}
		n_2(x)\,dx\right]
	\eeq
After integration, this becomes
	\beq
	\label{eq:weldon}
	\Sigma(E_p,T)
	&=	\frac{1}{16\pi}\frac{|\mathcal{M}_{1\to 2,M}|^2}{\beta\,\n p}
		\left[
			\ln\frac{ e^{\beta(\cb +\ca)}				+\eta_1 }
			{ e^{\beta(\cb -\ca)}						+\eta_1 }
		-	\ln\frac{ e^{\beta(\cb +\ca)}e^{-\beta E_p}	+\eta_2 }
			{ e^{\beta(\cb -\ca)}e^{-\beta E_p}			+\eta_2 }
		\right]\;,
	\eeq
where
	\beg
	\eta_i\equiv
	\begin{cases}
	-1&\text{particle $i$ is a~boson}\\
	+1&\text{particle $i$ is a~fermion}
	\end{cases}\;,\\
	\ca\equiv\frac{\n p\,\sqrt{\lambda(m_1^2,m_2^2,M^2)}}{2M^2}\;,\qquad
	\cb\equiv\frac{m_1^2-m_2^2+M^2}{2M^2}\,E_p\;,\\
	\lambda(m_1^2,m_2^2,M^2)\equiv\left[m_1^2-(m_2-M)^2\right]\left[m_1^2-(m_2+M)^2\right]\;,\\
	\beta\equiv \frac{1}{T}\;,\qquad
	|\vec p|\equiv\sqrt{E_p^2-M^2}\;.
	\eeg
This result differs slightly from \cref{eq:selfEFinal} with $X_0$ provided in \cref{app:X0table-VFDM}, namely, it lacks the~factor of $\nicefrac12$ or $\nicefrac13$ for a fermion or vector mediator, respectively.

\bibliography{references}
\bibliographystyle{JHEP}

\end{document}